\documentclass[pra,twocolumn,showpacs,preprintnumbers]{revtex4}
\usepackage[pdftex]{graphicx}
\graphicspath{{figures/pdf/},{figures/eps/}}
\usepackage{dcolumn}
\usepackage{bm}
\usepackage{amsmath} 
\usepackage{amssymb}  

\def\vec#1{\mathbf{#1}}

\def\ket#1{| #1 \rangle}
\def\bra#1{\langle #1 |}

\def\norm#1{|| #1 ||}
\def\sx{\sigma_x}
\def\sy{\sigma_y}
\def\sz{\sigma_z}

\def\D{\mathcal{D}}

\def\diag{\mbox{\rm diag}}

\def\Tr{\mathop{\rm Tr}}
\begin{document}
\DeclareGraphicsExtensions{.pdf}
\title{Effective Strategies for Identifying Model Parameters
for Open Quantum Systems}
\author{Er-ling Gong$^{1}$}
\author{Weiwei Zhou$^{1}$}
\author{S. G. Schirmer$^{2}$}
\author{Zhi-Qiang Sun$^{1}$}
\author{Ming Zhang$^{1}$}
\affiliation{
$^{1}$%
Department of Automatic Control, College of Mechatronic Engineering
and Automation, National University of Defense Technology\\
Changsha, Hunan 410073,  People's Republic of China\\
$^{2}$%
Department of Applied Mathematics and Theoretical Physics,
University of Cambridge, Cambridge, CB3 0WA, UK
}%
\date{\today}

\begin{abstract}
The problem of identifiability of model parameters for open quantum
systems is considered by investigating two-level dephasing systems.  We
discuss under which conditions full information about the Hamiltonian
and dephasing parameters can be obtained.  Using simulated experiments
several different strategies for extracting model parameters from
limited and noisy data are compared.
\end{abstract}

\maketitle

\section{Introduction}

Control and optimization of quantum systems have been recognized as
important issues for many years~\cite{Blaquiere1} and control theory for
quantum systems has been developed since the 1980s~\cite{Huang2,Ong3,
Clark4}.  There has been considerable recent progress in both theory and
experiment~\cite{qcb,R1-2000}.  However, despite this progress, there
are still many challenges.  Most quantum control schemes rely on
open-loop control design based on mathematical models of the system to
be controlled.  However, accurate models are not often not available,
especially for manufactured quantum systems such as artificial quantum
dot atoms or molecules.  Therefore, system identification~\cite{19} is a
crucial prerequisite for quantum control.

In the quantum information domain, procedures for characterization of
quantum dynamical maps are often known as quantum-process tomography
(QPT)~\cite{2,3,4} and many schemes have been proposed to identify the
unitary (or completely positive) processes, for example, standard
quantum-process tomography (SQPT)~\cite{1,5,6}, ancilla-assisted process
tomography (AAPT)~\cite{8,9,10} and direct characterization of quantum
dynamics (DCQD)~\cite{11}.  However, if control of the system's dynamics
is the objective, what we really need to characterize is not a global
process but the generators of the dynamical evolution such as the
Hamiltonian and dissipation operators.  The problem of Hamiltonian
tomography (HT), though less well-understood, has also begun to be
considered recently by a few authors~\cite{15,16,17,18}.  Although QPT
and HT differ in various regards, both try to infer information about
the quantum dynamics from experiments performed on systems, and both can
be studied from the point of view of system identification with broad
tasks including (1) experimental design and data gathering, (2) choice
of model sets and model calculation, and (3) model validation.

Recently the quantum system identification problem has been briefly
explored from cybernetical point of view, and underlining the important
role of experimental design~\cite{CDC}.  In this article we follow this
line of inquiry.  Throughout the paper, we make the following basic
assumptions: (1) the quantum system can be repeatedly initialized in a
(limited) set of known states; (2) that we can let the system evolve for
a desired time $t$; and (3) that some projective measurements can be
performed on the quantum system.  The main question we are interested in
in this context is how the choice of the initialization and measurement
affect the amount of information we can acquire about the dynamics of
the system.  Given any a limited range of options for the experimental
design, e.g., a range of measurements we could perform, different
choices for the initial states, or different control Hamiltonians, how
to choose the best experimental design, and what are the theoretical
limitations?  Finally, we are interested in efficient ways to extracting
the relevant information from noisy experimental data.

The paper is organized as follows: In Sec. II we discuss the model and
basic design assumptions.  Sec III deals with the general question of
model identifiability in various settings, and in Sec IV we compare
several different stategies for parameter estimation from a limited set
of noisy data from simulated experiments see how they measure up.

\section{Model and Design Assumptions}

To keep the analysis tractable we consider a simple model of a qubit
subject to a Hamiltonian $H$ and a system-bath interaction modelled by a
single Lindblad operator $V$, i.e., with system dynamics governed by the
master equation
\begin{equation}
 \label{3}
 \tfrac{\partial\rho(t)}{\partial{t}}
 =-\tfrac{i}{\hbar}[\hat{H},\rho]+\D[V](\rho),
\end{equation}
where the Lindbladian dissipation term is given by
\begin{equation}
 \label{4}
  \D[V](\rho) = V \rho V^\dag - \tfrac{1}{2}(V^\dag V + V V^\dag).
\end{equation}
We shall further simplify the problem by assuming that $V$ is a
Hermitian operator representing a depolarizing channel or pure phase
relaxation in some basis.  Without loss of generality we can choose the
basis so that $V$ is diagonal, in fact we can choose
$V=\sqrt{\tfrac{\gamma}{2}}\sz$ with $\sz=\diag(1,-1)$ and $\gamma\ge0$.
Under these assumptions the master equation simplifies
\begin{equation}
  \D[\sz](\rho) = \tfrac{\gamma}{2}(\sz\rho\sz-\rho).
\end{equation}
The control Hamiltonian can be expanded with respect to the Pauli basis
$\{\sx,\sy,\sz\}$
\begin{equation}
\label{5}
  H(t)=\tfrac{\hbar}{2}(\omega_{0}(t) \sz +\omega_{1}(t)\sx-\omega_{2}(t)\sy)
\end{equation}
with possibly time-dependent coefficients $\omega_\alpha(t)$.  It is
convenient to consider a real representation of the system.  Following
the approach in \cite{20} we expand $\rho$ with respect to the
standard Pauli basis for the $2\times 2$ Hermitian matrices
\begin{equation}
\label{6-2} \rho=\tfrac{1}{2}(I+v_x \sx + v_y \sy + v_z\sz),
\end{equation}
where the coefficients are $v_{\alpha}=\Tr(\rho\sigma_{\alpha})$.
Similarly expanding the dynamical operators allows us to recast
Eq.~(\ref{3}) in following Bloch equation ($\hbar=1$)
\begin{equation}
\label{7}
\begin{pmatrix}
\dot{v}_{x}(t)\\ \dot{v}_{y}(t)\\ \dot{v}_{z}(t)
\end{pmatrix} =
\begin{pmatrix}
-\gamma&-\omega_{0}(t)&-\omega_{2}(t)\\
\omega_{0}(t)&-\gamma&-\omega_{1}(t)\\
\omega_{2}(t)&\omega_{1}(t)&0
\end{pmatrix}
\begin{pmatrix}
{v}_{x}(t)\\ {v}_{y}(t)\\ {v}_{z}(t)
\end{pmatrix}.
\end{equation}

Using this simple model for illustration we subsequently consider the
experimental design from three aspects: (1) initialization procedures,
(2) measurement choice and (3) Hamiltonian design.

\textbf{(1) Initialization.}  
We assume the ability to prepare the system in some initial state
\begin{equation}
 \label{IV}
 |\psi_{I}(0)\rangle
 = \cos\tfrac{\theta_{I}}{2}|0\rangle+\sin\tfrac{\theta_{I}}{2}|1\rangle,
\end{equation}
with respect to the basis $\{\ket{0},\ket{1}\}$, which coincides with
the eigenbasis of $V$.  We can formally represent the initialization
procedure by the operator $\Pi(\theta_I)$, which is the projector onto
the state $\ket{\psi_I}$, with $I$ indicating initialization.  With
these restrictions the design of the initialization procedure is reduced
to the selection of parameter $\theta_{I}$.  Note that we assume that we
can only prepare one fixed initial state, not a full set of basis
states.

\textbf{(2) Measurement.} 
We assume the ability to perform a two-outcome projective measurement
\begin{equation}
  \label{M}
  M = M_+ - M_- = \ket{m_+}\bra{m_+} - \ket{m_-}\bra{m_-},
\end{equation}
where the measurement basis states can be written as
\begin{subequations}
\label{m+-}
\begin{align}
 \ket{m_{+}}
 &= \cos\tfrac{\theta_{M}}{2}|0\rangle+\sin\tfrac{\theta_{M}}{2}|1\rangle \\
 \ket{m_{-}}
 &= \sin\tfrac{\theta_{M}}{2}|0\rangle-\cos\tfrac{\theta_{M}}{2}|1\rangle
\end{align}
\end{subequations}
so that the choice of the measurement can be reduced to suitable
choice of the parameter $\theta_M$, and we shall indicate this by
writing $M(\theta_M)$.

\textbf{(3) Hamiltonian.} In practice we may or may not have the freedom
to choose the type of Hamiltonian but it will be instructive to consider
the identification problem for the following three cases:
\begin{itemize}
\item[(a)] $\omega_{z}(t)\equiv{\omega_{z}}>0$ and
           $\omega_{x}(t)=\omega_{y}(t)\equiv0$.

\item[(b)] $\omega_{x}(t)\equiv{\omega_{x}}>0$ and
           $\omega_{y}(t)=\omega_{z}(t)\equiv0$.

\item[(c)] $\omega_{y}(t)\equiv{\omega_{y}}>0$ and
           $\omega_{x}(t)=\omega_{z}(t)\equiv0$.
\end{itemize}

In this context the experimental design problem can be reduced to the
problem of choosing suitable values $\theta_{I}$ and $\theta_{M}$ to
identify the system parameters $\gamma$ and $\omega_*$ in each of the
three cases above.  The problem considered here is similar to that
considered in \cite{18}, in particular we still only allow a single
initial state and single fixed measurement, but unlike in \cite{18} the
initial state and the measurement are not assumed to commute with the
dephasing operator.

\section{Model Identifiability}

We now consider the identification problem for the dephasing qubit
system, concentrating on \emph{experimental design} issues for
simultaneously discriminating dephasing parameter and Hamiltonian
parameter of two-level dephasing systems.

\subsection{$H=\omega_z\sz$, $V=\sqrt{\tfrac{\gamma}{2}}\sz$}

In this special case $H$ commutes with the dephasing operator $V$.
Solving the equation
\begin{equation}
\label{5-7b}
\begin{pmatrix}
 \dot{v}_{x}(t)\\ \dot{v}_{y}(t)\\ \dot{v}_{z}(t)
\end{pmatrix} =
\begin{pmatrix}
  -\gamma&-\omega_{z}&0\\ \omega_{z}&-\gamma&0\\ 0&0&0
 \end{pmatrix}
\begin{pmatrix}
 {v}_{x}(t)\\ {v}_{y}(t)\\ {v}_{z}(t)
\end{pmatrix}
\end{equation}
with the initial state (\ref{IV}) gives
\begin{equation}
\label{5-8b}
\begin{pmatrix}
{v}_{x}(t)\\
{v}_{y}(t)\\
{v}_{z}(t)
\end{pmatrix} =
\begin{pmatrix}
e^{-\gamma{t}}\cos\omega_{z}t\sin\theta_{I}\\
e^{-\gamma{t}}\sin\omega_{z}t\sin\theta_{I}\\
\cos\theta_{I}
\end{pmatrix}
\end{equation}
and applying the binary-outcome projective measurement $M(\theta_M)$
yields the measurement traces $p_{\pm}(t)=\Tr[M_{\pm}\rho(t)]$.
Defining $\bar{p}_{\pm}(t) = 2p_{\pm}(t)-1$, we have
\begin{equation}
\label{5-10b}
  \bar{p}_{\pm}(t)
  = \pm\cos\theta_{I}\cos\theta_{M}
    {\pm}e^{-\gamma{t}}\cos\omega_{z}t\sin\theta_{I}\sin\theta_{M},
\end{equation}
from which we can see that we can obtain information about the system
parameters $\omega_z$ and $\gamma$ if and only if $\sin\theta_I \neq 0$
and $\sin\theta_M \neq 0$, i.e., if neither the initial state
preparation $\Pi(\theta_I)$ nor the measurement $M$ commutes with $H$
and $V$.  The measurement traces also yield information about $\theta_I$
and $\theta_M$, i.e., we can determine the relative angles between the
initialization and measurement axis and the fixed Hamiltonian/dephasing
axis, if they are not known a-priori.  We also see that the visibility
is maximized if $\sin\theta_I\sin\theta_M=1$, which will be the case if
the initialization and measurement axis are orthogonal to the joint
Hamiltonian and dephasing axis.

These results make physical sense.  As $[H,V]=0$, if the initial state
preparation $\Pi(\theta_I)$ commutes with $H$ and $V$ then the initial
state is a stationary state of the dynamics and the measurement outcome
is constant in time $c_{\pm}=\tfrac{1}{2}(1\pm 1)$.  If
$\sin\theta_I\neq 0$ then the initial state is not stationary and the
state follows a spiral path towards the joint Hamiltonian and dephasing
axis but as both $H$ and $V$ are proportional to $\sz$, $\Tr[\sz
\rho(t)]$ is a conserved quantity of the dynamics.  If $\sin\theta_M=0$
then the measurement commutes with $\sz$, and as $\Tr[\sz\rho(t)]$ is a
conserved quantity, we again obtain no information about the dynamics
from which to identify the system parameters.

\begin{figure*}
\scalebox{0.45}{\includegraphics[viewport=0 0 561 420]{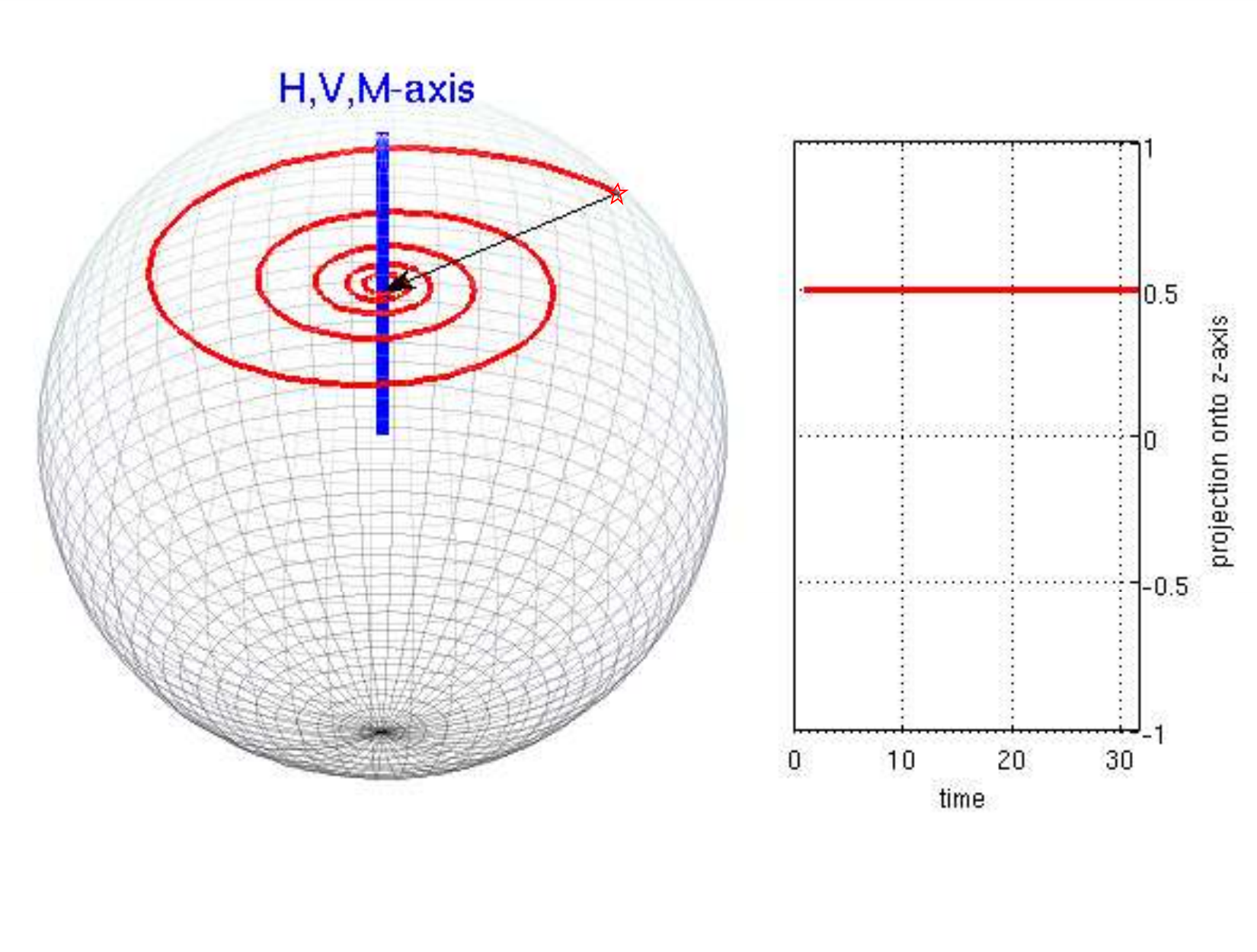}}
\scalebox{0.45}{\includegraphics[viewport=0 0 561 420]{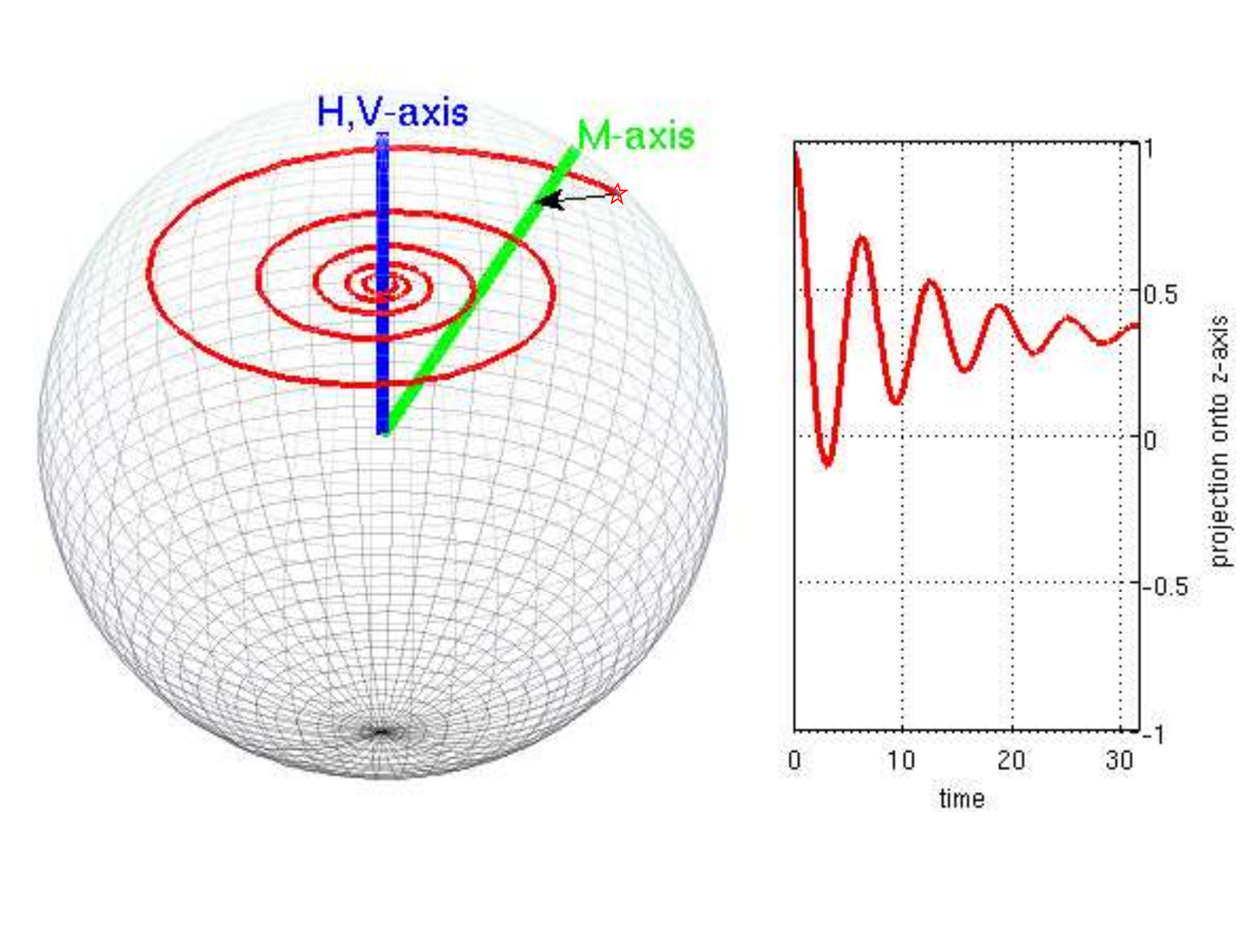}}
\caption{Evolution of system state on the Bloch sphere and projection
onto measurement axis for $H=V$: In the pathological case when the
measurement axis coincides with the $H,V$-axis, no information about the
system parameters can be obtained (left); Otherwise the measurement
trace contains information about both $H$ and $V$ (right), provided the
initial state is not stationary.}  \label{fig:bloch1}
\end{figure*}

\subsection{$H=\omega_x\sx$, $V=\sqrt{\tfrac{\gamma}{2}}\sz$}

In this case, the equation~(\ref{7}) is reduced into the following
equation
\begin{equation}
\label{5-7c} \left(\begin{array}{c}
\dot{v}_{x}(t)\\
\dot{v}_{y}(t)\\
\dot{v}_{z}(t)
\end{array}\right)=\left(\begin{array}{ccc}
-\gamma&0&0\\
0&-\gamma&-\omega_{x}\\
0&\omega_{x}&0
\end{array}\right)\left(\begin{array}{c}
{v}_{x}(t)\\
{v}_{y}(t)\\
{v}_{z}(t)
\end{array}\right)
\end{equation}
and the solution of Eq.~(\ref{5-7c}) for the initial state
(\ref{IV}) is
\begin{equation}
\label{5-8c}
\begin{pmatrix}
{v}_{x}(t)\\
{v}_{y}(t)\\
{v}_{z}(t)
\end{pmatrix}
=\begin{pmatrix}
e^{-\gamma{t}}\sin\theta_{I}\\
\Phi^{x}_{2}(t)\cos\theta_{I}\\
\Phi^{x}_{3}(t)\cos\theta_{I}
\end{pmatrix}
\end{equation}
where we set
$\widehat{\omega}_{x}=\sqrt{\omega_{x}^{2}-\tfrac{\gamma^{2}}{4}}$ and
\begin{subequations}
\begin{align}
\Phi^{x}_{2}(t)
 &=-e^{-\tfrac{\gamma}{2}{t}}
  \tfrac{\omega_{x}}{\widehat{\omega}_x}\sin\widehat{\omega}_{x}t \label{a123}\\
\Phi^{x}_{3}(t)
 &=e^{-\tfrac{\gamma}{2}{t}}[\cos\widehat{\omega}_{x}t
   +\tfrac{\gamma}{2\widehat{\omega}_{x}}\sin\widehat{\omega}_{x}t].\label{a133}
\end{align}
\end{subequations}
If $\omega_x^2<\gamma^2/4$ then $\widehat{\omega}$ will be purely
imaginary and the sine and cosine terms above are replaced by the
respective hyperbolic functions.  If $\omega_x^2=\gamma^2/4$, the
expression $\widehat{\omega}^{-1}\sin(\widehat{\omega} t)$ must be
analytically continued.

Eq.~(\ref{5-8c}) implies that we can obtain the estimated value of the
following probabilities
\begin{equation}
\label{5-10c}
 \bar{p}_{\pm}(t)
 =  \pm e^{-\gamma{t}}\sin\theta_{I}\sin\theta_{M}
    \pm\Phi^{x}_{3}(t)\cos\theta_{I}\cos\theta_{M}.
\end{equation}
Noting that $\Phi_{3}^x(t)$ depends on both $\omega_x$ and $\gamma$ this
shows that we can obtain full information about the system parameters if
and only if $\cos\theta_{I}\neq 0$ and $\cos\theta_{M}\neq 0$.  If
$\cos\theta_{I}=0$ or $\cos\theta_{M}=0$ then we can still identify
$\gamma$ but not $\omega_x$.  Again this makes sense.

$\cos\theta_I=0$ for $\theta_I=\tfrac{\pi}{2}$, i.e., if the initial
state is an eigenstate of the Hamiltonian.  Since $[H,V]\neq 0$ in this
case, eigenstates of $H$ are not stationary.  However, since the
Hamiltonian and dephasing axis are \emph{orthogonal}, the initial state
remains in a plane orthogonal to the dephasing axis, the $z=0$ plane in
our case, following the path $x(t)=e^{-\gamma t}$.  Thus we have
$[H,\rho(t)]=0$ for all times, and we can therefore not obtain any
information about the Hamiltonian parameter $\omega_x$, but we can still
obtain information about the dephasing parameter $\gamma$.  If the
Hamiltonian and dephasing axis were not orthogonal then we would expect
to be able to identify both the Hamiltonian and dephasing parameters
even if the initial state was an eigenstate of $H$ as in this case it
would not remain an eigenstate of $H$ under the evolution.

If $\cos\theta_I \neq 0$ but $\cos\theta_M=0$ then the measurement
commutes with the Hamiltonian.  Transforming to the Heisenberg picture,
\begin{equation*}
  \dot{M}_{\pm}(t)
 = -i [M_{\pm}(t),H] -\tfrac{\gamma}{2} \D[\sz]M_{\pm}(t) ,
\end{equation*}
and again one can show that $M_{\pm}(t)$ remains orthogonal to the
dephasing axis and $\Tr[M_{\pm}(t)\rho_0]=\Tr[M_{\pm}\rho(t)]$ is
independent of the Hamiltonian $H$, explaining why we cannot obtain any
information about $H$ in this case.

\begin{figure*}
\scalebox{0.45}{\includegraphics[viewport=0 0 561 420]{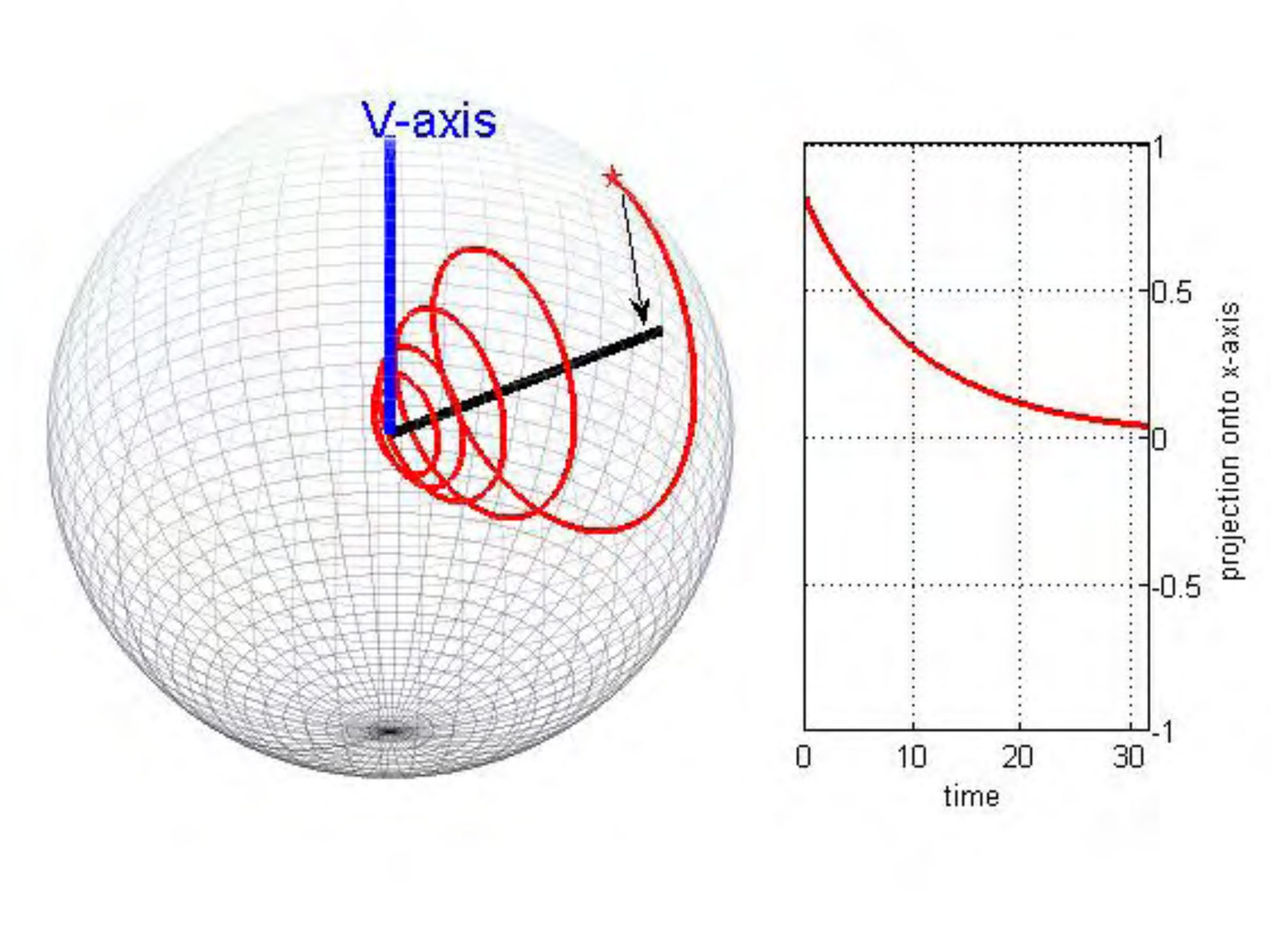}}
\scalebox{0.45}{\includegraphics[viewport=0 0 561 420]{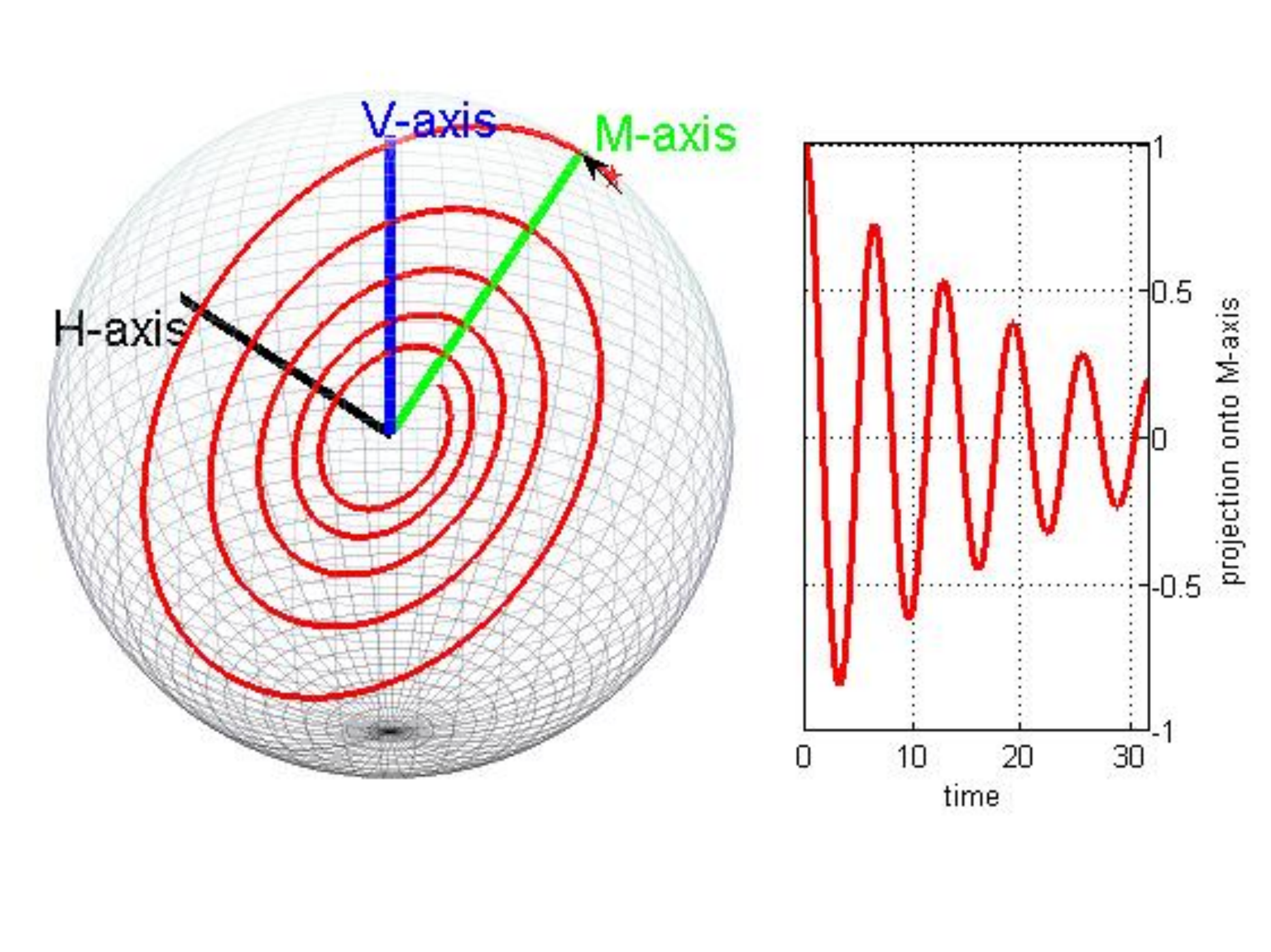}}
\caption{Evolution of system state on the Bloch sphere and projection
onto measurement axis for $H\perp V$: If the measurement axis coincides
with the $H$-axis, only information about the decoherence parameter
$\gamma$ can be obtained (left).  In case $C$ the measurement trace
always contains information about both $H$ and $V$ (right) for any
initial state and measurement.}  \label{fig:bloch2}
\end{figure*}

\subsection{$H=\omega_y\sy$, $V=\sqrt{\tfrac{\gamma}{2}}\sz$}

In this case equation (\ref{7}) is reduced into the following equation
\begin{equation}
\label{5-11} \begin{pmatrix}
\dot{v}_{x}(t)\\
\dot{v}_{y}(t)\\
\dot{v}_{z}(t)
\end{pmatrix}
=\begin{pmatrix}
-\gamma&0&-\omega_{y}\\
0&-\gamma&0\\
\omega_{y}&0&0
\end{pmatrix}
\begin{pmatrix}
{v}_{x}(t)\\
{v}_{y}(t)\\
{v}_{z}(t)
\end{pmatrix}
\end{equation}
whose solution for the initial state (\ref{IV}) is
\begin{equation}
\label{5-10m-1a}
\begin{pmatrix}
{v}_{x}(t)\\
{v}_{y}(t)\\
{v}_{z}(t)
\end{pmatrix}
= \begin{pmatrix}
 \Phi_{23}^y(t)\sin\theta_{I} -e^{-\tfrac{\gamma}{2}{t}}
 \tfrac{\omega_{y}}{\widehat{\omega}_{y}}\sin\widehat{\omega}_{y}t
 \cos\theta_{I}\\
  0\\
 \Phi_{33}^y(t)\cos\theta_{I} +
  e^{-\tfrac{\gamma}{2}{t}} \tfrac{\omega_{y}}{\widehat{\omega}_{y}}
\sin\widehat{\omega}_{y}t\sin\theta_{I}
\end{pmatrix}
\end{equation}
where
$\widehat{\omega}_{y}=\sqrt{\omega_{y}^{2}-\tfrac{\gamma^{2}}{4}}$ and
\begin{subequations}
\begin{align}
 \Phi_{23}^y(t) &= e^{-\tfrac{\gamma}{2}{t}}
 [\cos\widehat{\omega}_yt-
  \tfrac{\gamma}{2\widehat{\omega}_{y}}\sin\widehat{\omega}_{y}t]\\
 \Phi_{33}^y(t) &= e^{-\tfrac{\gamma}{2}{t}}
 [\cos\widehat{\omega}_{y}t+\tfrac{\gamma}{2\widehat{\omega}_{y}}
  \sin\widehat{\omega}_{y}t]
\end{align}
\end{subequations}
This implies that  we can obtain the estimated value of the probabilities
\begin{equation}
\label{5-10d}
\bar{p}_{\pm}(t)
 = {\pm} \alpha_1 e^{-\tfrac{\gamma}{2}{t}} \cos\widehat{\omega}_{y}t \\
   \pm \alpha_2 e^{-\tfrac{\gamma}{2}{t}} \sin\widehat{\omega}_{y}t
\end{equation}
where the coefficient functions are
\begin{subequations}
\label{eq:modelC:coeff}
\begin{align}
 \alpha_1 &= \cos(\theta_{I}-\theta_{M})\\
 \alpha_2 &= \tfrac{\gamma}{2\widehat{\omega}_{y}}\cos(\theta_{I}+\theta_{M})
   +\tfrac{\omega_{y}}{\widehat{\omega}_{y}}\sin(\theta_{I}-\theta_{M}).
 \end{align}
\end{subequations}
As before, if $\omega_y^2<\gamma^2/4$ then $\widehat{\omega}$ will be
purely imaginary and the sine and cosine terms above turn into their
respective hyperbolic sine and cosine equivalents, and if
$\omega_y^2=\gamma^2/4$, the expression $\widehat{\omega}^{-1}
\sin(\widehat{\omega} t)$ must be analytically continued.

In this case it is quite interesting to notice that it is impossible to
find such $\theta_{I}$ and $\theta_{M}$ that
$\cos(\theta_{I}-\theta_{M})
=\sin(\theta_{I}-\theta_{M})=\cos(\theta_{I}+\theta_{M})=0$ and thus
that we can identify both model parameters for any choice of the initial
state and measurement.  This also makes sense because regardless of the
choice of $\theta_I$ and $\theta_M$ the initial state in this case is
always orthgonal to the Hamiltonian axis, and the measurement $M$ is
always orthogonal to $H$, $[M,H]\neq 0$, there are no conserved
quantities and the only stationary state of the system is the completely
mixed state.

\section{Parameter Estimation}

In this section we explore how to estimate dephasing and Hamiltonian
parameters from limited noisy measurement data.  In an actual experiment
we can only estimate the probabilities $p_\pm(t)$ at a finite number of
times $t_k$ by repeatedly initializing the system in some fixed state
$\rho_0$, and letting it evolve for time $t_k$ before performing the
projective measurement $M$.  Each single repetition of the experiment
yields a binary outcome $+1$ or $-1$ and we can estimate the probability
$p_{\pm}(t_k)$ by repeating the experiment $N_e$ times and computing the
relative frequencies of the respective measurement outcomes $\pm 1$,
e.g., $\widehat{p}_\pm(t_k)=\tfrac{N_\pm}{N_{e}}$.

To model noisy experimental data we could generate
$\widehat{p}_{\pm}(t)$ by adding a zero-mean White-Gaussian noise signal
$\widehat{g_{t}}$ to $p_{\pm}(t)$, i.e., $\widehat{p}_\pm(t)= p_\pm(t)+
\widehat{g_{t}}$.  By the Law of Large Numbers and Iterated-logarithm
Law~\cite{dong1} this gives a Gaussian distribution $\widehat{p}_\pm(t)$
with mean $p_{\pm}(t)$ and variance $\sigma^2\sim \tfrac{\log\log N_e}{2
N_e}$ for $N_e\rightarrow\infty$.  For $N_e$ large this should be a good
error model, but for small $N_e$ it may not accurately capture the nature
of the projection noise, which follows a Poisson distribution.  To more
accurately model noisy experimental data when the number of measurement
repetitions is relatively small we can simulate the actual experiment by
generating $N_e$ random numbers $r_n$ between $0$ and $1$, drawn from a
uniform distribution, and setting $\widehat{p}_+=N_0/N_e$, where $N_0$ is
the number of $r_n\le p_+$.

\subsection{Fourier and time-series analysis}

\begin{figure*}
\includegraphics[width=\columnwidth]{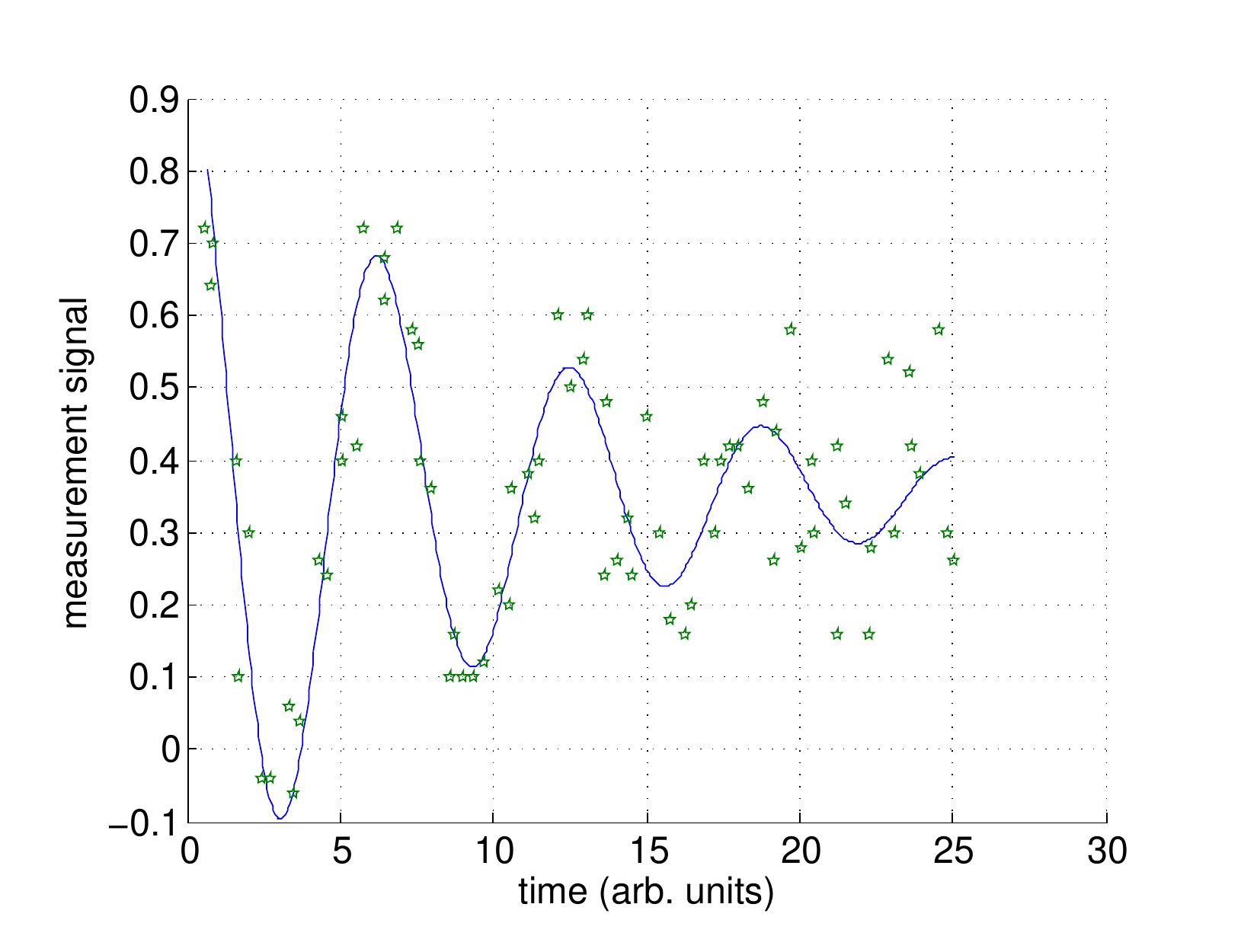}
\includegraphics[width=\columnwidth]{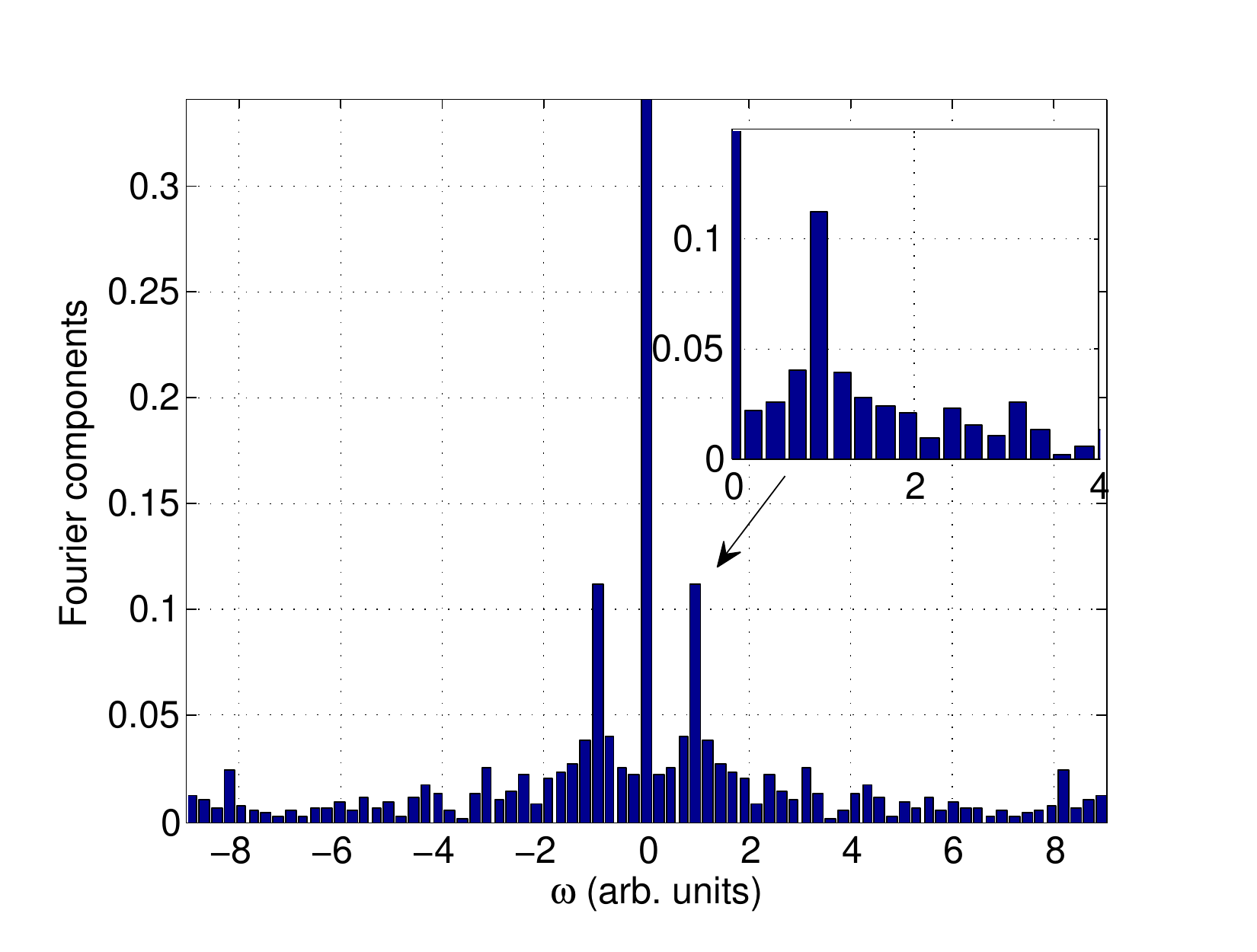}
\caption{Ideal signal and sparse noisy data points for a simulated
experiment assuming Model A with $\omega=1$ and $\gamma=0.1$ (left).  We
can still identify a peak in the Fourier spectrum (right) around
$\omega=1$ but the accuracy of this estimate is limited, and estimating
$\gamma$ from the peak broadening would be a challenge.}
\label{fig:exp1}
\end{figure*}

One commonly used technique to find frequency components in a noisy
time-domain signal is the Fourier transform or its discrete version, the
discrete or Fast Fourier Transform (FFT).  In principle, this allows us
to estimate the Hamiltonian parameters (frequencies) $\omega_{z}$,
$\hat{\omega}_{x}$ and $\hat{\omega}_{y}$~\cite{17}, and the damping
rate $\gamma$ can be estimated from the Lorentzian broadening of the
Fourier peak~\cite{18}.  When the signal is sparse, i.e., we only have a
relatively small number of sample points, and noisy, however, this
approach becomes problematic.  Fig.~\ref{fig:exp1} shows that we can
still identify a peak in the spectrum for sparse noisy signals but the
accuracy of the estimate is limited, and accurately estimating the
dephasing rate $\gamma$ from the broadening of the peak, given the
distortion, is very challenging.

Alternatively, once an estimated value for $\omega_{z}$ has been
obtained, $\gamma$ can be deduced in other ways.  In the simplest case A,
where the evolution is given by Eq.~(\ref{5-10b}), we have
\begin{equation}
  e^{-\gamma t} \cos\omega _z t \sin \theta _I \sin \theta _M =
  \bar{p}_+(t) - \cos\theta _I\cos\theta _M
\end{equation}
which we can rewrite as
\begin{equation}
  \gamma t = -\log \left(
  \tfrac{\bar{p}_+(t) - \cos\theta _I\cos\theta _M}
        {\cos\omega _z t\sin \theta _I \sin \theta _M} \right)
  \buildrel \Delta\over = z(t).
\end{equation}
Given $\bar{p}_+(t)$, if $\theta_I$, $\theta_M$ and $\omega_z$ are
known, we can therefore in principle calculate $z(t)$ and thus
$\hat{\gamma}$.  However, in practice the problem is that $p_{+}(t)$ and
all the other parameters are not known precisely.  In this case we could
estimate $\gamma$ using the Least Square Method~\cite{zhang1} and the
principle of Sequential Analysis\cite{zhang2,zhang3}.

\begin{figure*}
\includegraphics[width=\columnwidth]{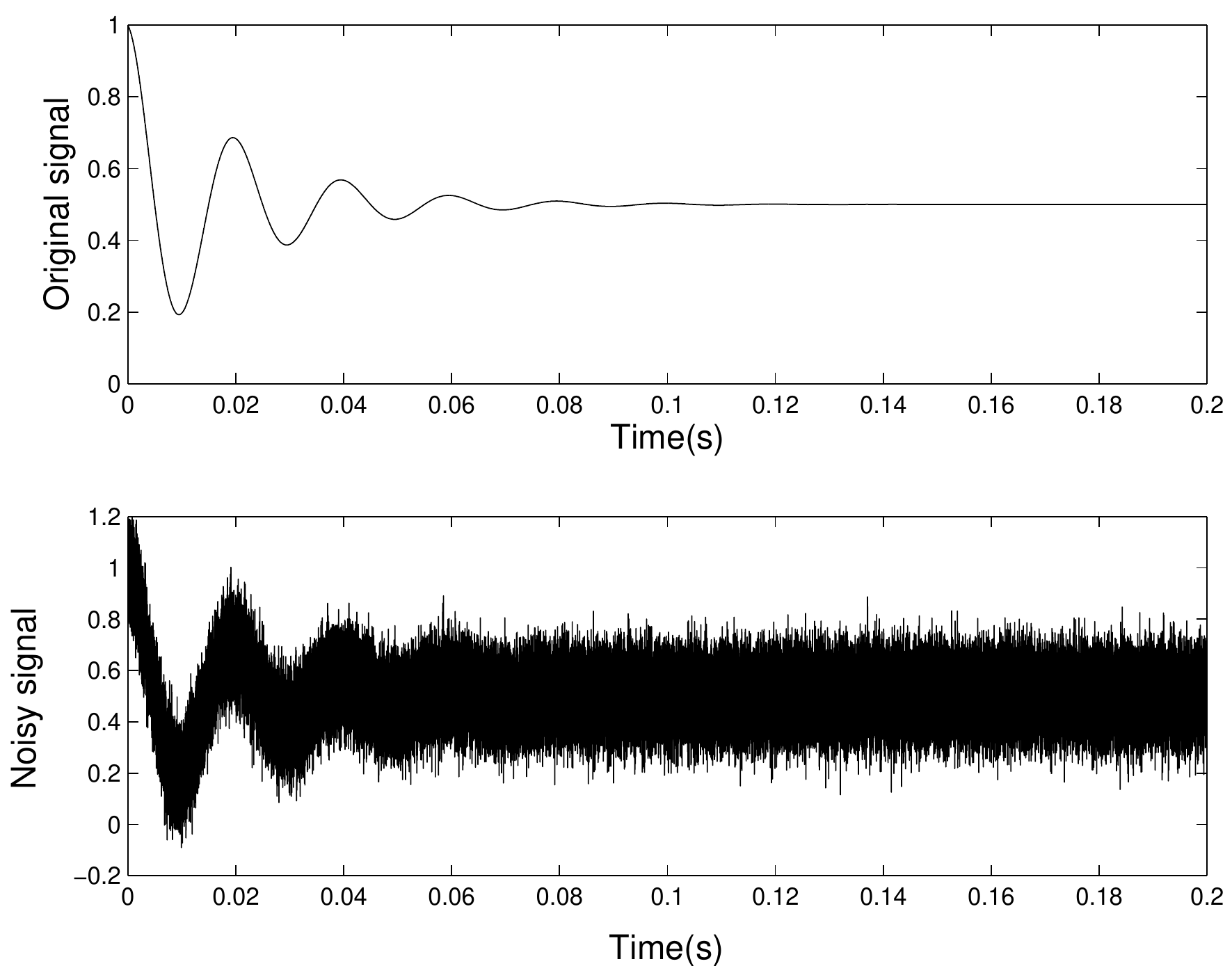}
\includegraphics[width=\columnwidth]{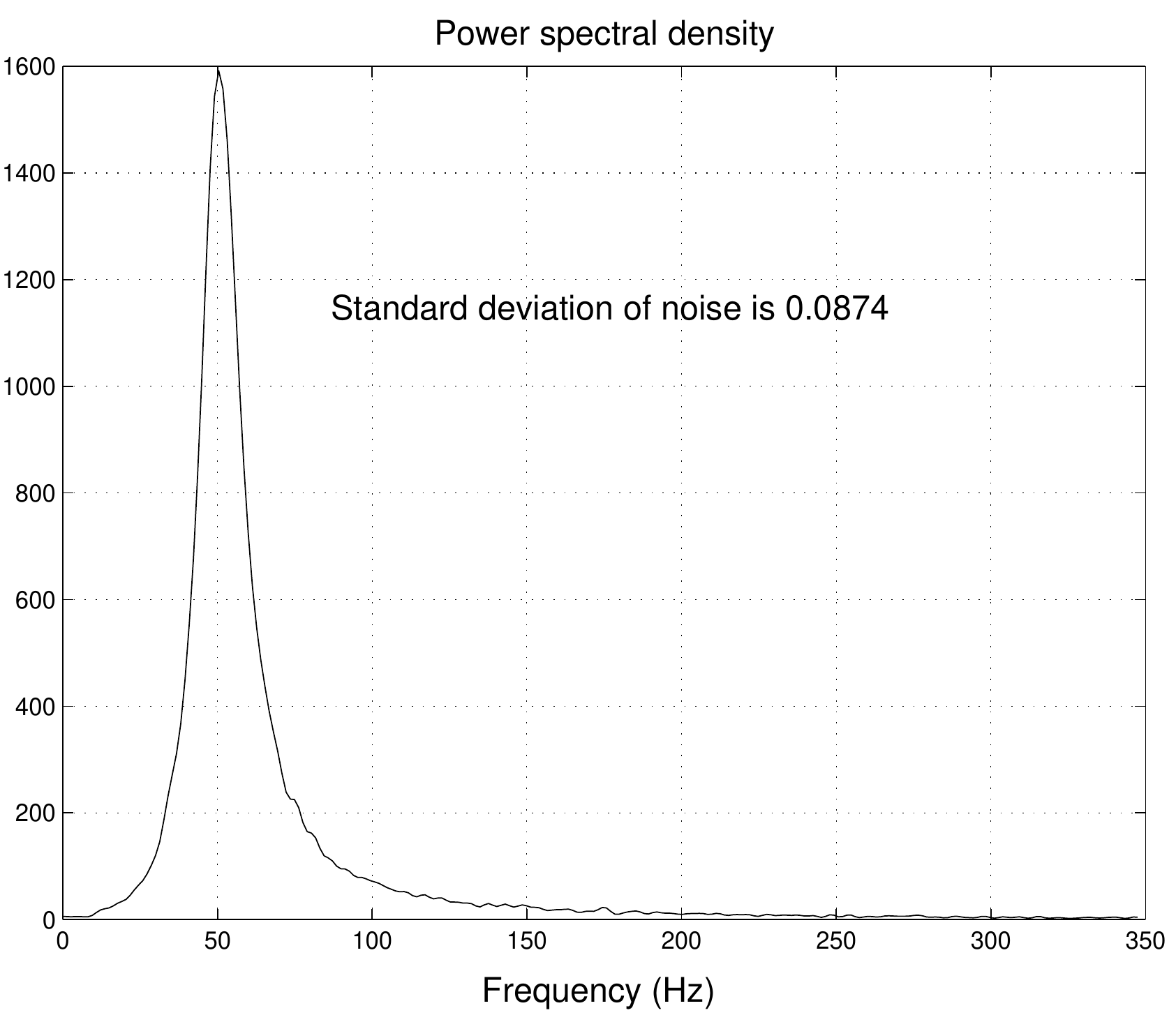}
\caption{Original time domain signal and simulated noisy signal (left)
and power spectral density of noisy signal (right).}  \label{figp}
\end{figure*}
\begin{figure*}
\includegraphics[width=\columnwidth]{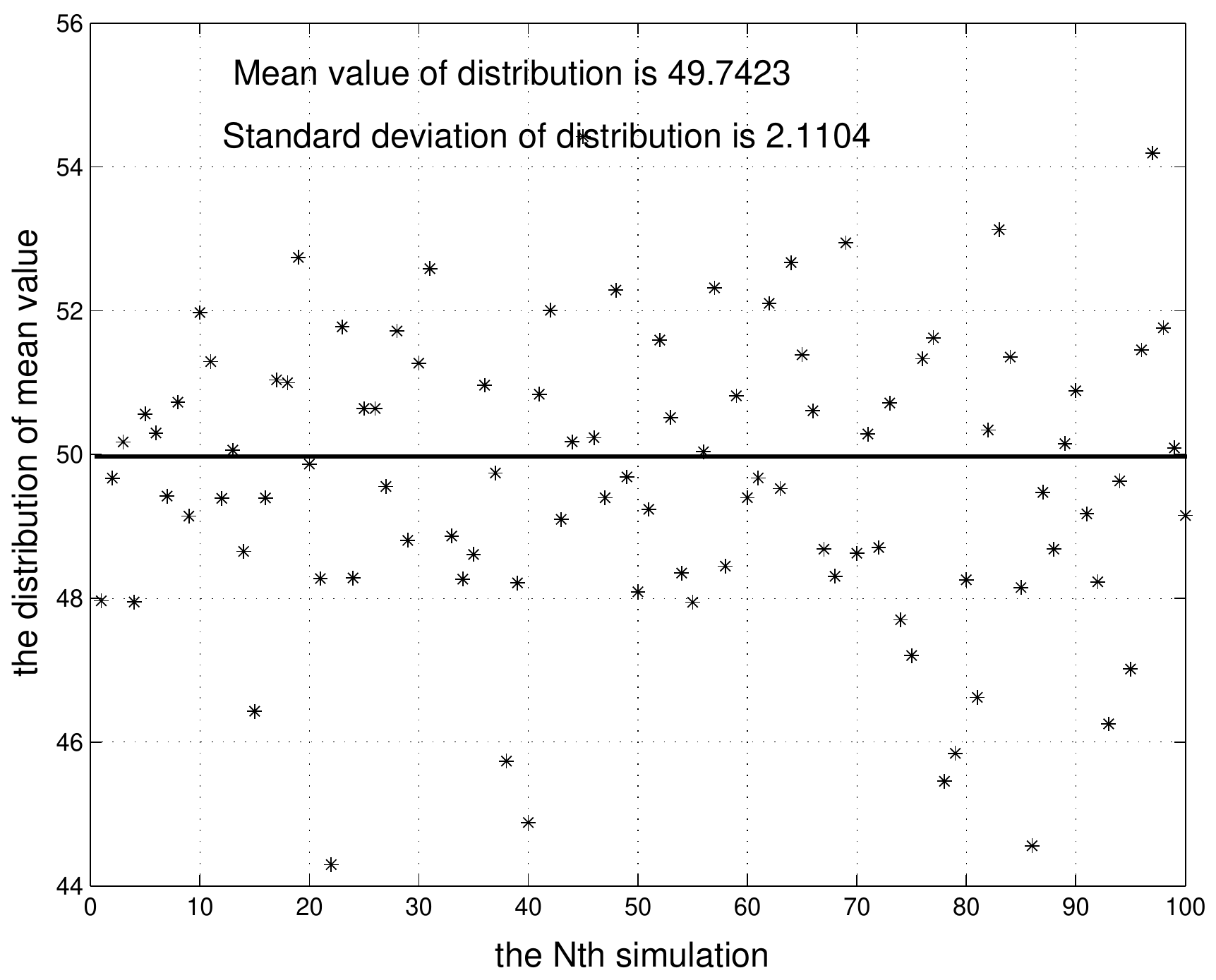}
\includegraphics[width=\columnwidth]{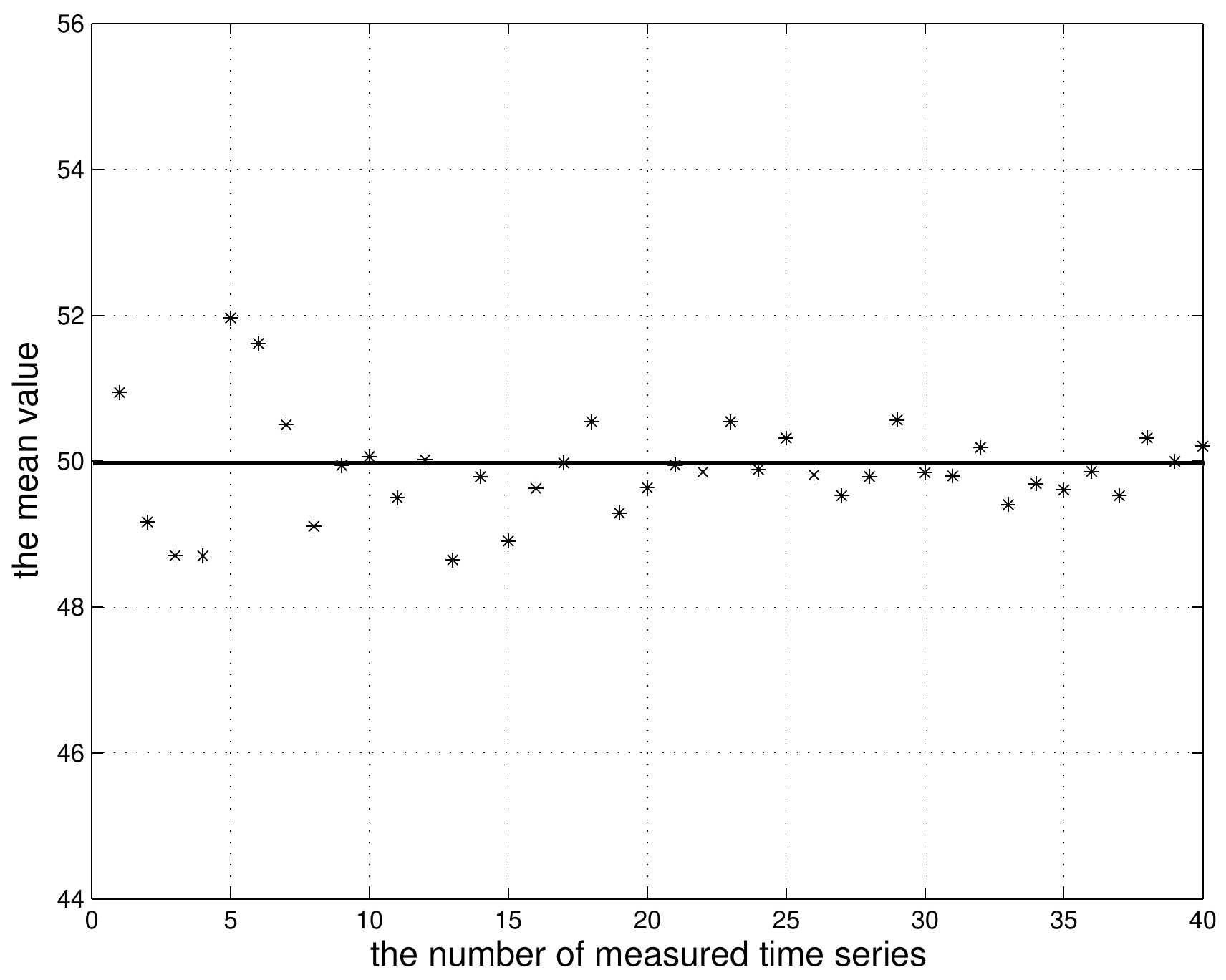}
\caption{Distribution of $\hat{\gamma}$ based on one simulated time
series (left) and mean value of $\hat{\gamma}$ for $1-40$ simulated
time series (right).}\label{figm}
\end{figure*}

To explore this approach we generated simulated noisy data signals
$\widehat{p}_+(t)$ with standard deviation $\sigma_{t}$ according to the
procedure above, assuming $\theta_{I}=\theta_{M}=\tfrac{\pi}{2}$,
$\omega_{z}=50$, $\gamma=50$ and the sampled period $T_{s}=1.4\mu$s.  If
the number of measurements $N_e$ at each time $t$ is $N_e=100$ then the
standard deviation of estimated error is $\sqrt{\tfrac{\log\log N_e}{2
N_e}}=0.0874$.  The original noiseless probability $\bar{p}_+(t)$, the
noisy signal with standard deviation $0.0874$ and the power spectral
density are shown in Fig.\ref{figp}.  Again the frequency $\omega_{z}$
can be easily determined by the peak value in Fig.~\ref{figp}(right).
We attempt to determine $\hat{\gamma}$ based on $\widehat{p}_+(t)$ using
simulated time-series.  For a single time series $\widehat{p}_+(t)$ the
distribution of $\hat{\gamma}$ around the true value of $50$ is shown in
Fig.~\ref{figm}(left).  The estimation is unsatisfactory even if the
signal is sampled densely at a high time resolution.  The results can be
noticeably improved by averaging over multiple time-series of
$\widehat{p}_+(t)$.  If the mean value of the $\gamma$-estimates over
several time series is used as the final estimate for $\gamma$, the
estimated error of $\hat{\gamma}$ can be remarkably reduced as shown in
Fig.~\ref{figm} (right), which indicates the mean value of
$\hat{\gamma}$ with $1-40$ time-series respectively.

%
%

\begin{figure*}
\includegraphics[width=\columnwidth]{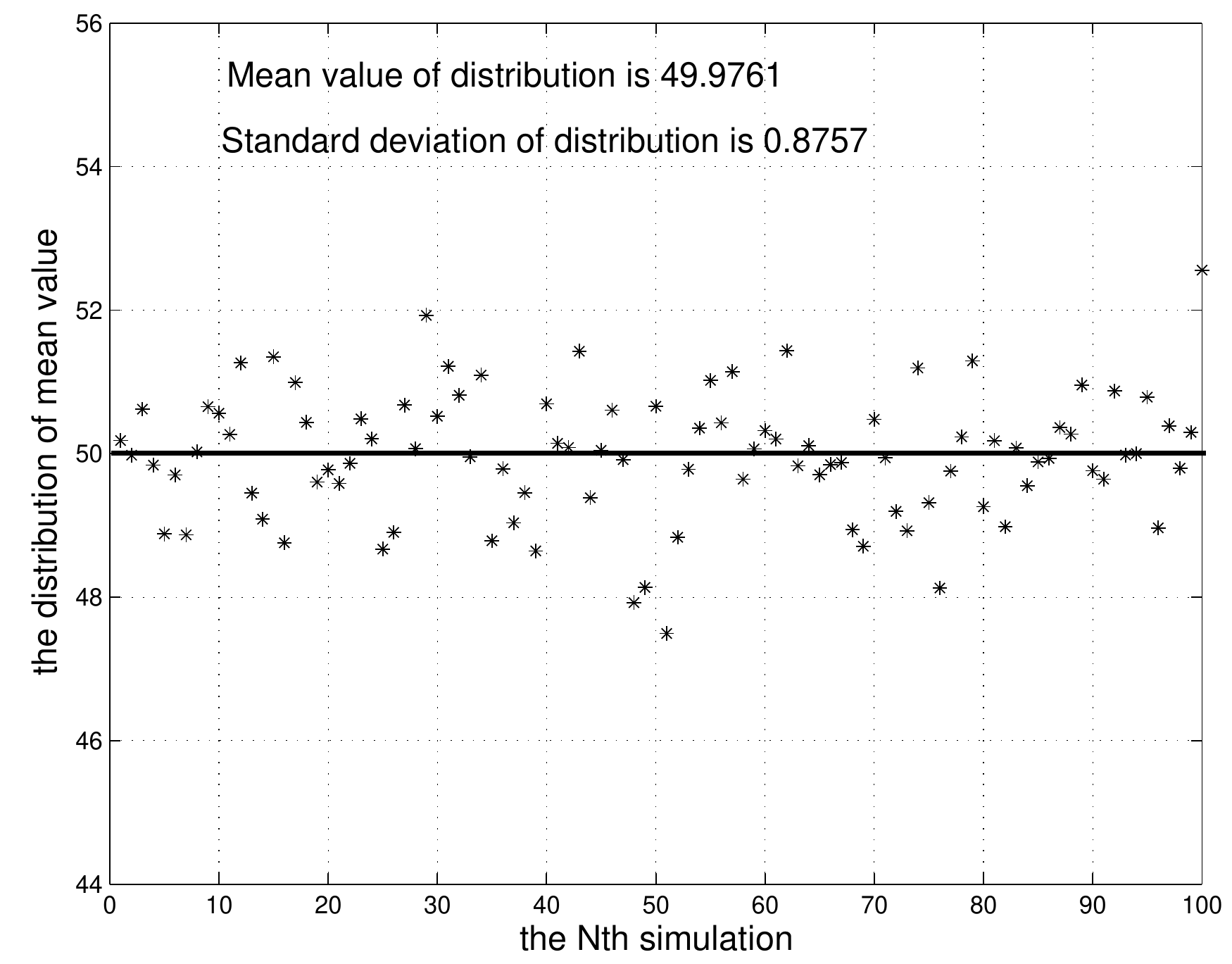}
\includegraphics[width=\columnwidth]{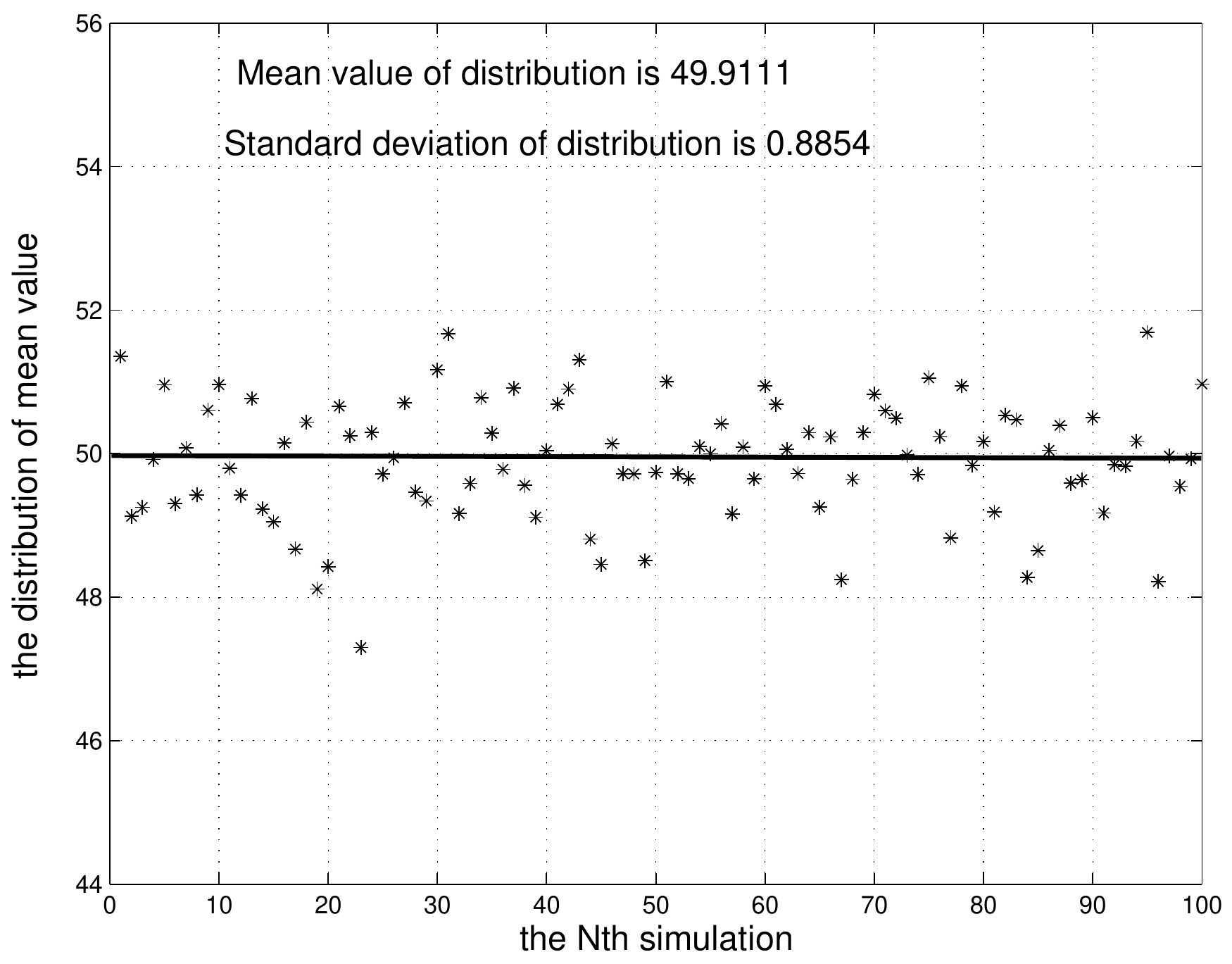} \caption{Distribution of mean
value of $\hat{\gamma}$ for $N_e=125$ using $4$ simulated time
series (left), and for $N_e=250$ using $2$ simulated time series
(right).} \label{figh}
\end{figure*}

The number of simulated time-series of $\widehat{p}_+(t)$ required
is determined by the desired estimation accuracy.  To achieve a mean
value of the distribution of ca.\ $50$ with standard deviation less
than $1.0$ simulations suggest that at least $5$ time-series with
$N_e=100$ are necessary, i.e., about $500$ measurements have to be
performed at each sample time $t$ to ensure that the standard
deviation of the estimated error is less than $1.0$.  Instead of
five time series with $N_e=100$ we could choose four times series
with $N_e=125$ or two with $N_e=250$ to reach the nearly same
estimation accuracy, as shown in Fig.~\ref{figh}.  In the actual
experiment one can obtain the mean-value sequence of $\hat{\gamma}$
by continually updating $p(t)$ at each sampled time $t$. If the
mean-value sequence is found to converge to a certain fixed value
and its standard deviation satisfies the requirement, the experiment
measurement should be stopped.


To estimate $\gamma$ by finite measurement data one will have to
identify a suitable time $t_{d}$ when to terminate the experiment,
ideally when the true value of $e^{-\gamma{t_{d}}}$ is almost $0$.  A
$p_+(t)$ is almost equal to $p_-(t)$ for $t\geq{t_{d}}$, the signal
after this time will be effectively a pure noise signal.  Therefore
increasing the signal length beyond a certain critical time $t_{k_{0}}$
will not improve the estimation accuracy.

\begin{figure}
\includegraphics[width=\columnwidth]{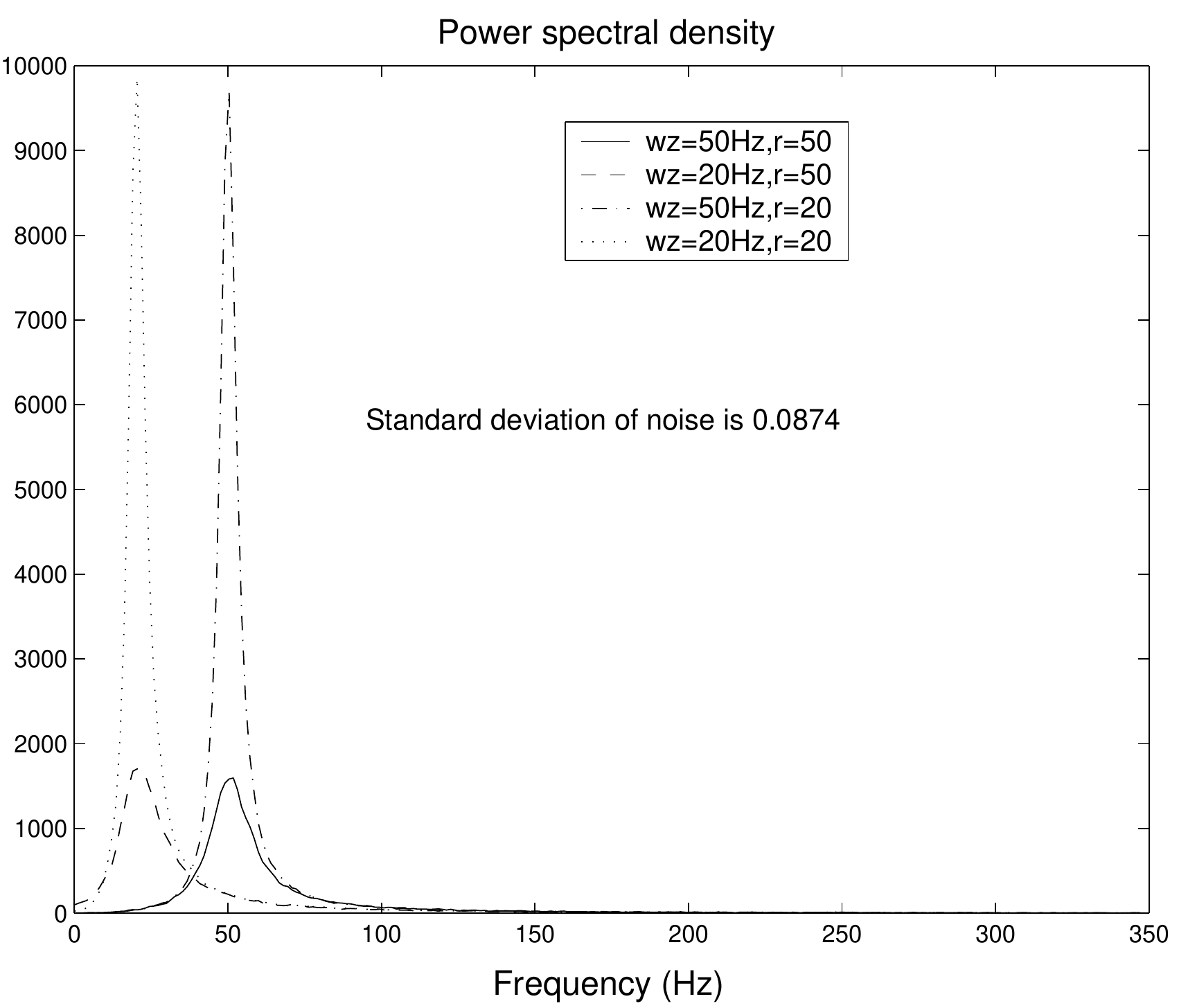}
\caption{Power spectral density of measured signal in different
cases} \label{figa}
\end{figure}

\begin{figure*}
\includegraphics[width=\columnwidth]{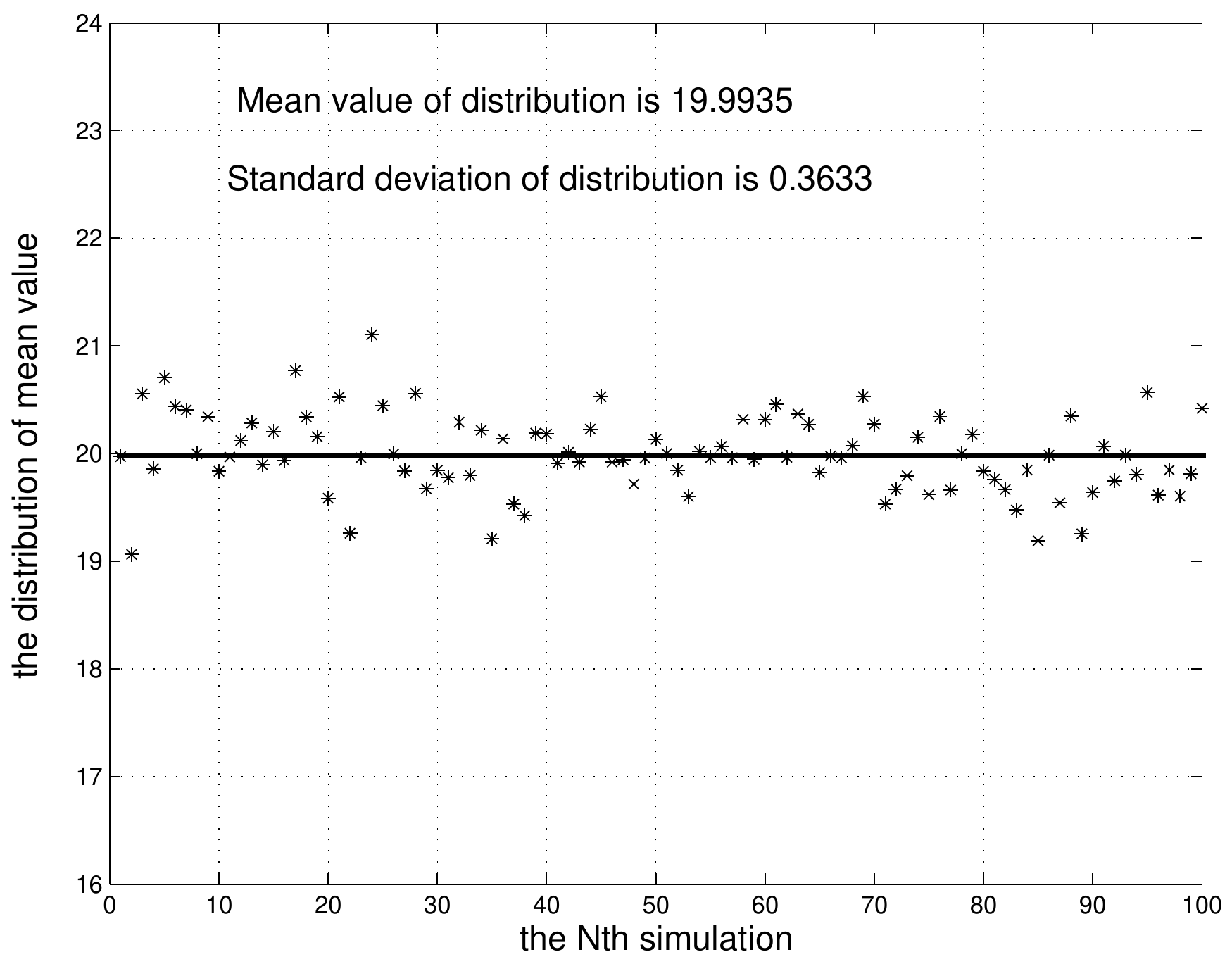}
\includegraphics[width=\columnwidth]{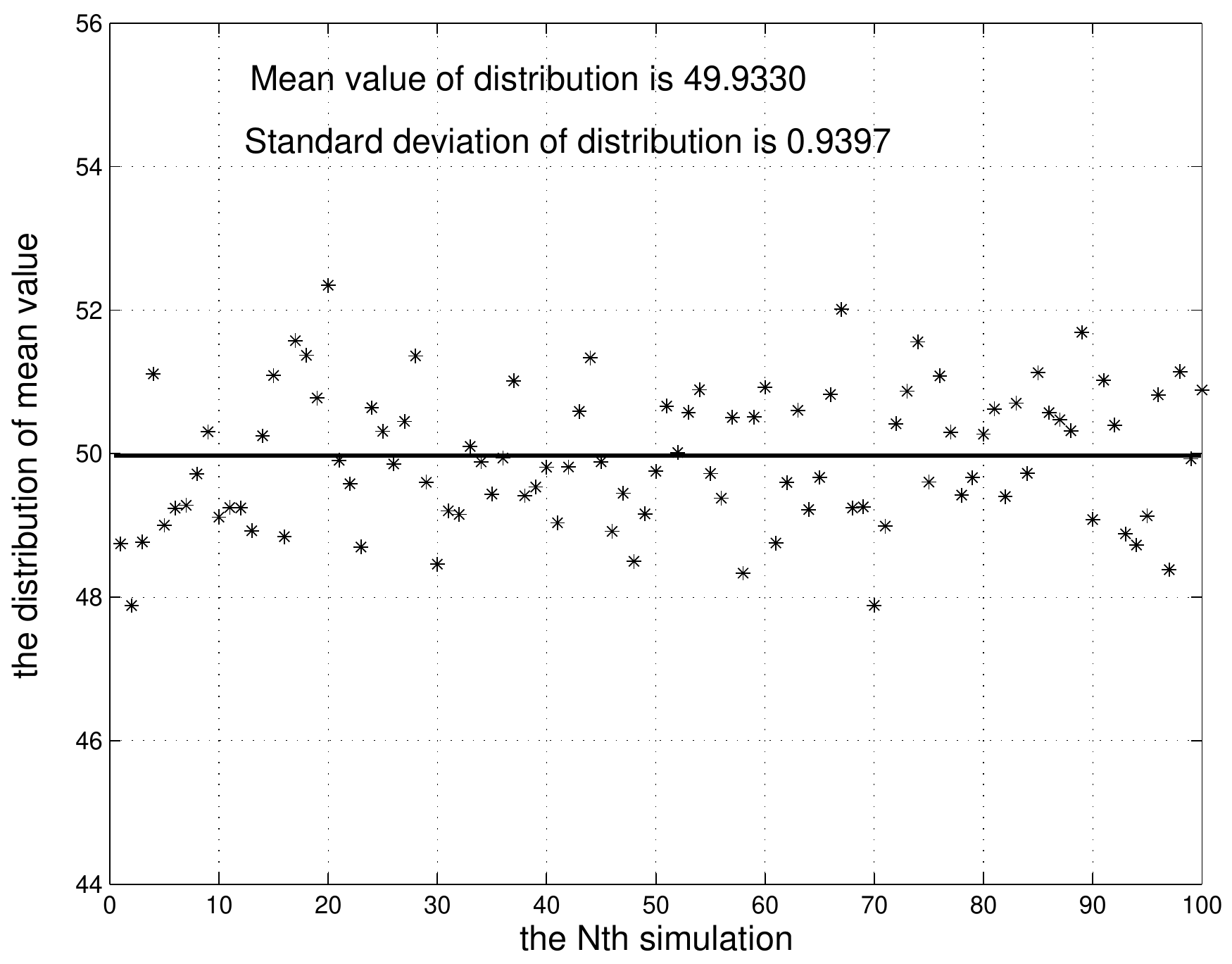}
\caption{Distribution of mean value of $\hat{\gamma}$ for
$\omega_{z}=50Hz$, $\gamma=20$ (left) and $\omega_{z}=20Hz$,
$\gamma=50$ (right).} \label{fige}
\end{figure*}

Furthermore, we consider the effect of different $\omega_{z}$ and
$\gamma$ on the estimation accuracy.  If the true value of $\gamma$ is
less than the value of $50$ above, the estimation of $\omega_{z}$ and
$\gamma$ will be more accurate (as shown in Fig.~\ref{figa}) as the
exponential $e^{-\gamma t}$ decays more slowly.  If the value of
$\omega_{z}$ is reduced, i.e., the oscillation period is increased while
its envelope remains unchanged the accuracy of the estimation is not
influenced (as shown in Fig.~\ref{fige}).

\subsection{Bayesian Analysis of Sparse Signals}

The previous section shows that we can in principle simultaneously
identify both dephasing parameter and Hamiltonian parameters for simple
open systems using Fourier and time-series analysis.  Both Fourier and
time-series analysis can deal with very noisy signals but they require
rather dense sampling of the signal with is both time-consuming and
resource-intensive.  Ideally we would like to be able to estimate the
parameters by sparse sampling the signal.  One promising approach in
this regard is Bayesian estimation.  The main idea behind the Bayesian
approach is to choose the parameters to be determined, here $\omega$ and
$\gamma$, to maximize a certain likelihood function
\begin{equation}
 L(\vec{p},\vec{d},\sigma) =
  \sigma^{-N}\exp\left[-\frac{\norm{\vec{p}-\vec{d}}_2^2}{2\sigma^2}\right].
\end{equation}
where $\vec{d}$ is the measured data and the $\vec{p}$ are the
probabilities predicted by the model, which depend on the parameters
to be determined, and $\sigma$ is the error variance.

Following the same approach as in \cite{16}, we write the signals as a
linear combination a small number of basis functions determined by the
functional form of the signals.  Here the measurement signals
$\bar{p}_{\pm}(t)$ can be written as a linear combination of two basis
functions
\begin{equation}
  \bar{p}_{\pm}(t) = \alpha_1 g_1(t) + \alpha_2 g_2(t).
\end{equation}
The values of the basis functions and the coefficients are given in
Table~\ref{table1}


\begin{table}
\begin{tabular}{|l|c|c|c|}
\hline
           & Model A & Model B & Model C \\\hline
$g_1(t)$   & $1$
           & $e^{-\gamma t}$
           & $e^{-\gamma t/2}\cos(\widehat{\omega}t)$ \\
$g_2(t)$   & $e^{-\gamma t}\cos(\omega_z t)$
           & $e^{-\gamma t/2} s(t)$
           & $e^{-\gamma t/2}\sin(\widehat{\omega}t)$ \\
$\alpha_1$ & $\cos(\theta_I)\cos(\theta_M)$
           & $\sin(\theta_I)\sin(\theta_M)$
           & (\ref{eq:modelC:coeff}a) \\
$\alpha_2$ & $\sin(\theta_I)\sin(\theta_M)$
           & $\cos(\theta_I)\cos(\theta_M)$
           & (\ref{eq:modelC:coeff}b) \\\hline
\end{tabular}
\caption{Chosen basis functions and coefficients for Bayesian
estimation. $s(t)=\cos(\widehat{\omega}t)+
\tfrac{\gamma}{2\widehat{\omega}}\sin(\widehat{\omega}t)$.
$\widehat{\omega}=\sqrt{\omega^2-(\tfrac{\gamma}{2})^2}$.}  \label{table1}
\end{table}

\begin{table}
\begin{tabular}{|l|ccc|ccc|}\hline
         &    & $\alpha_1$ &        &     & $\alpha_1$ &  \\
         & act    & est    & uncert. & act & est & uncert. \\\hline
 Model A & 0.3536 & 0.3536 & 0.0133 & 0.6124 & 0.5267 & 0.1393 \\
 Model B & 0.6124 & 0.5952 & 0.0491 & 0.3536 & 0.3531 & 0.0514 \\
 Model C & 0.9659 & 1.0066 & 0.0594 & 0.2332 & 0.2506 & 0.0573\\\hline
\end{tabular}
\caption{Actual and estimated values of the linear coefficients
$\alpha_1$ and $\alpha_2$.}\label{table2}
\end{table}

\begin{figure*}
\includegraphics[width=\columnwidth]{Model1-SimExp}
\includegraphics[width=\columnwidth]{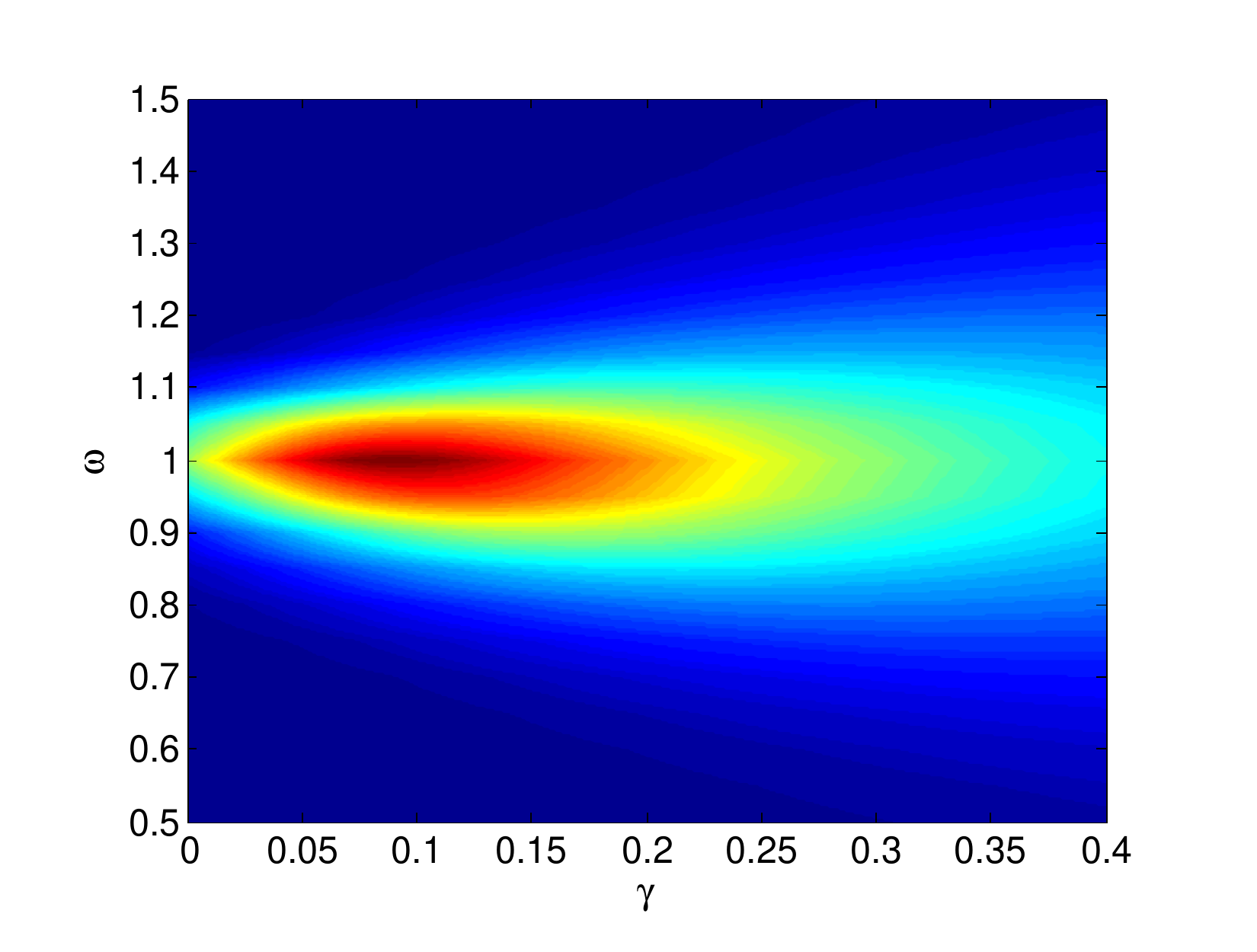}
\caption{Ideal signal and sparse noisy sampling ($N=75$ samples,
signal length $T\approx 25$, accuracy level $N_e=100$) (left) and
corresponding log-likelihood function (right) for model system A
with $\omega_z=1$ and $\gamma=0.1$.}  \label{fig:BayesianA}
\end{figure*}

\begin{figure*}
\includegraphics[width=\columnwidth]{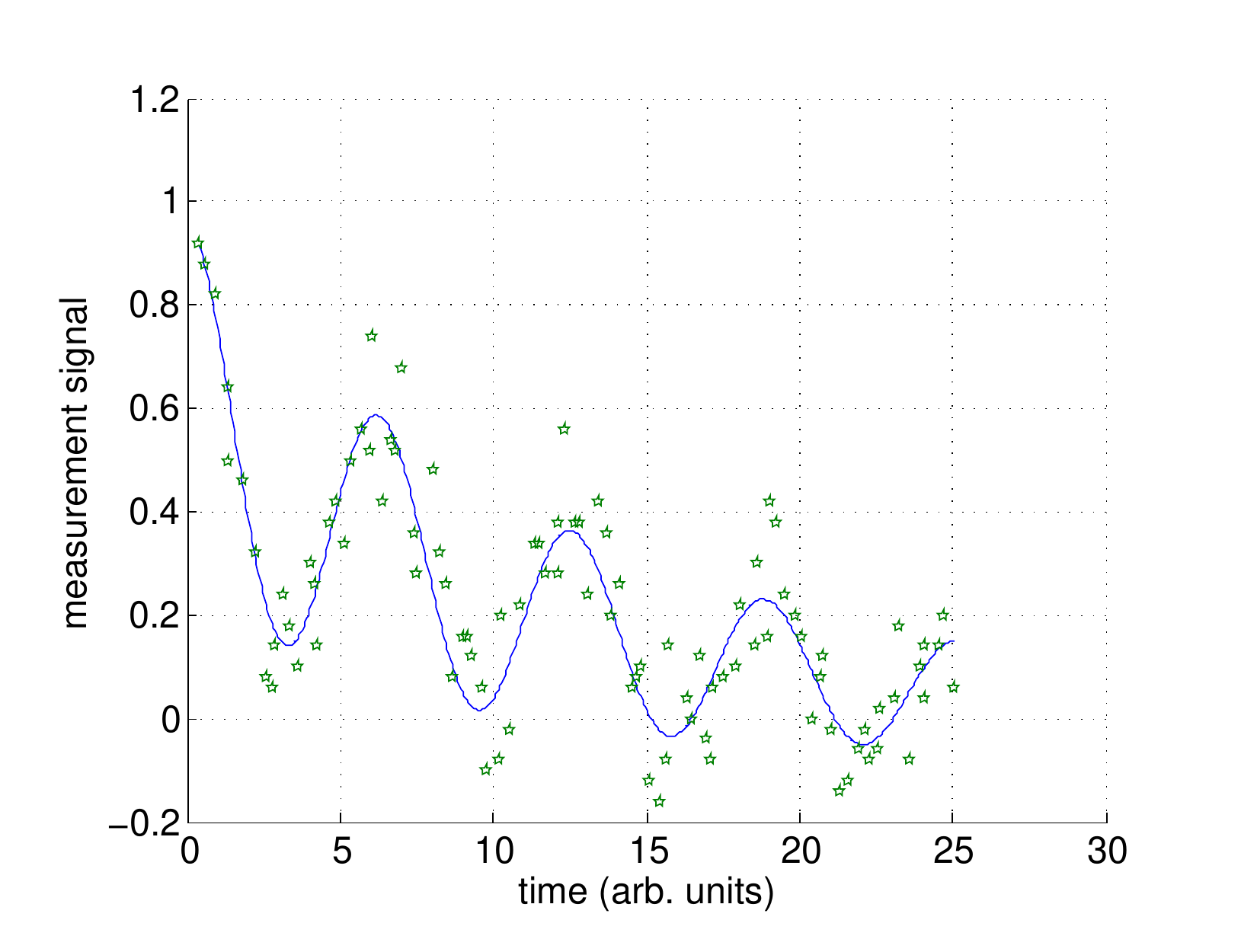}
\includegraphics[width=\columnwidth]{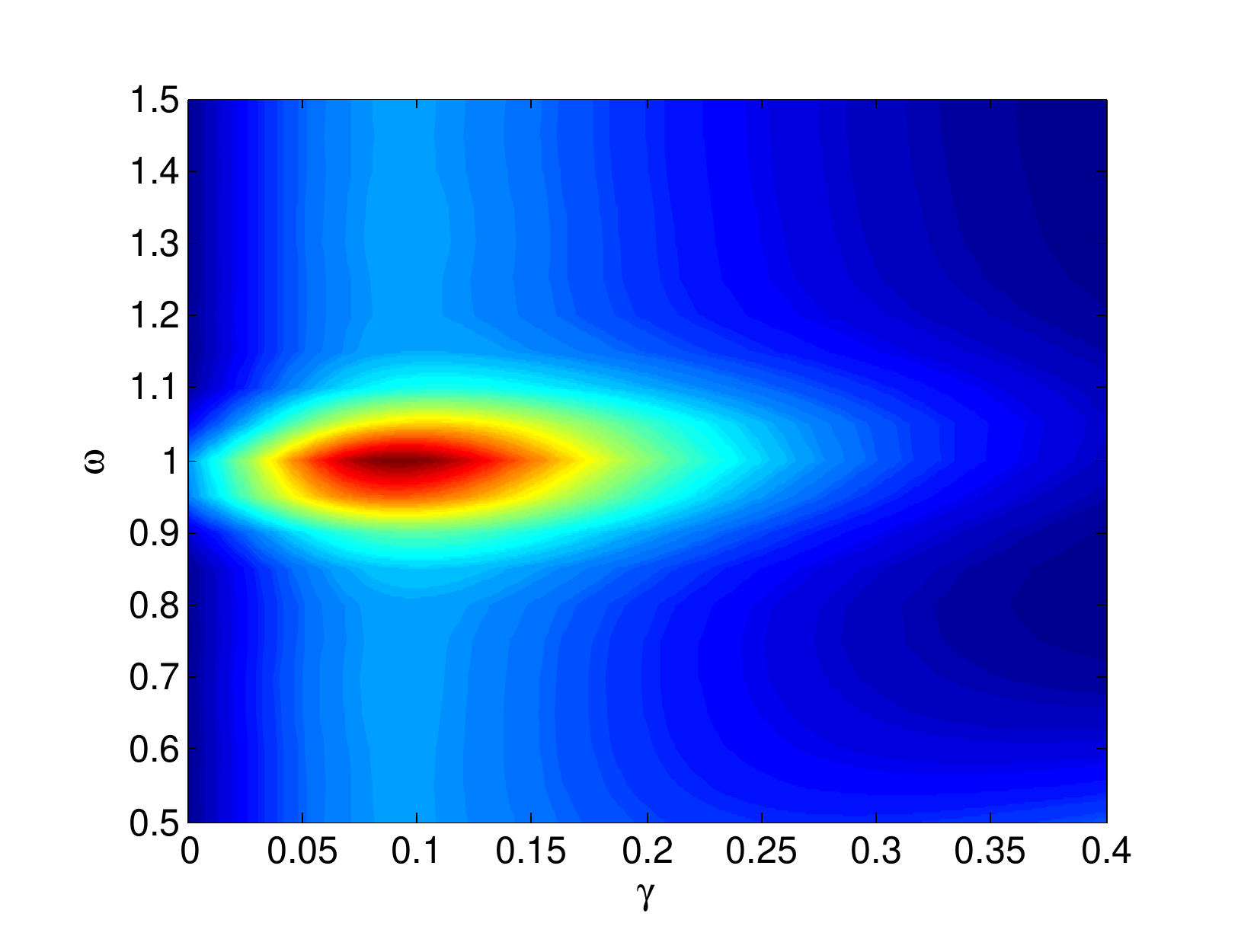}
\caption{Ideal signal and sparse noisy sampling ($N=100$, $T\approx 25$,
$N_e=100$) (left) and corresponding log-likelihood function (right) for
model system B with $\omega_x=1$ and $\gamma=0.1$.}  \label{fig:BayesianB}
\end{figure*}

\begin{figure*}
\includegraphics[width=\columnwidth]{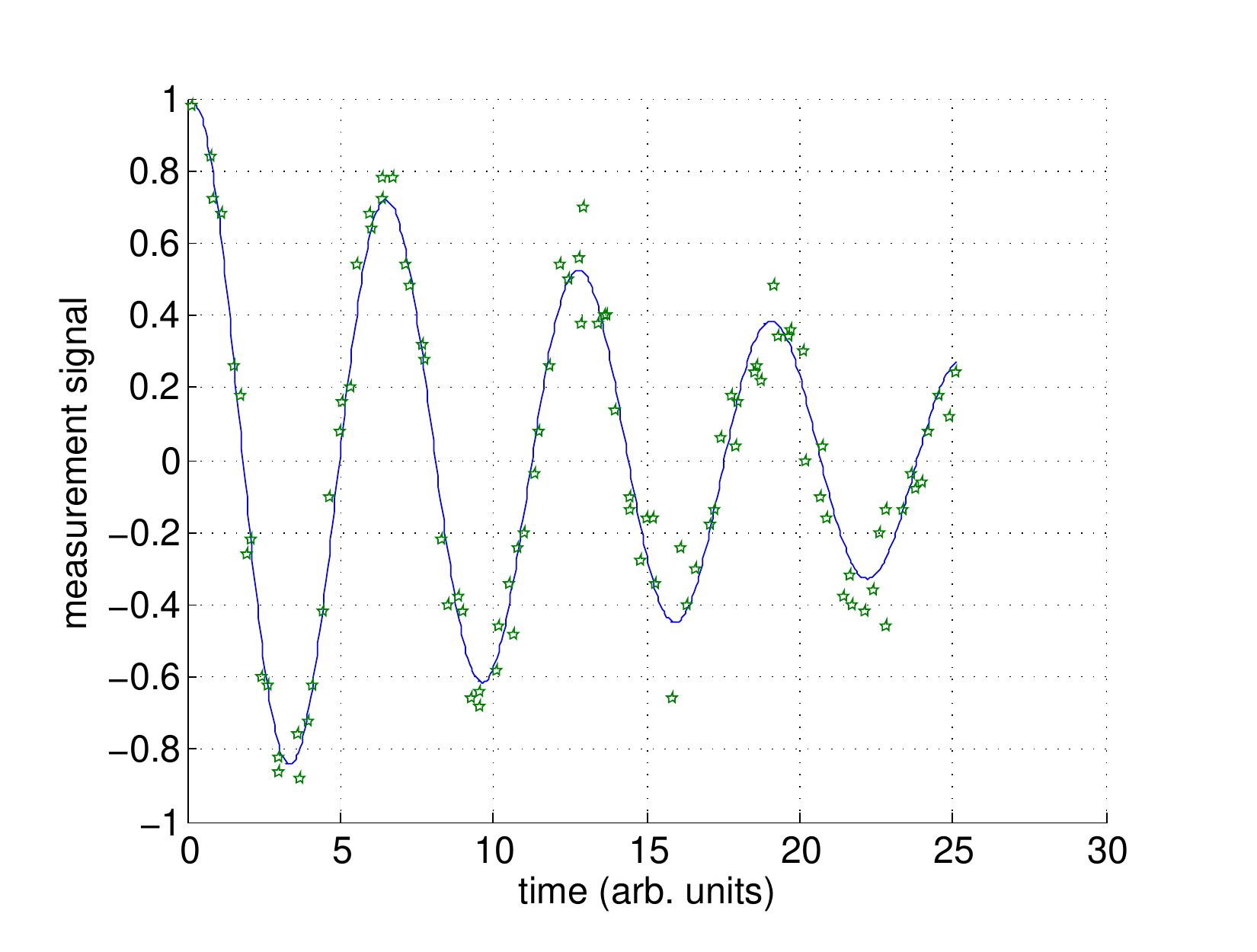}
\includegraphics[width=\columnwidth]{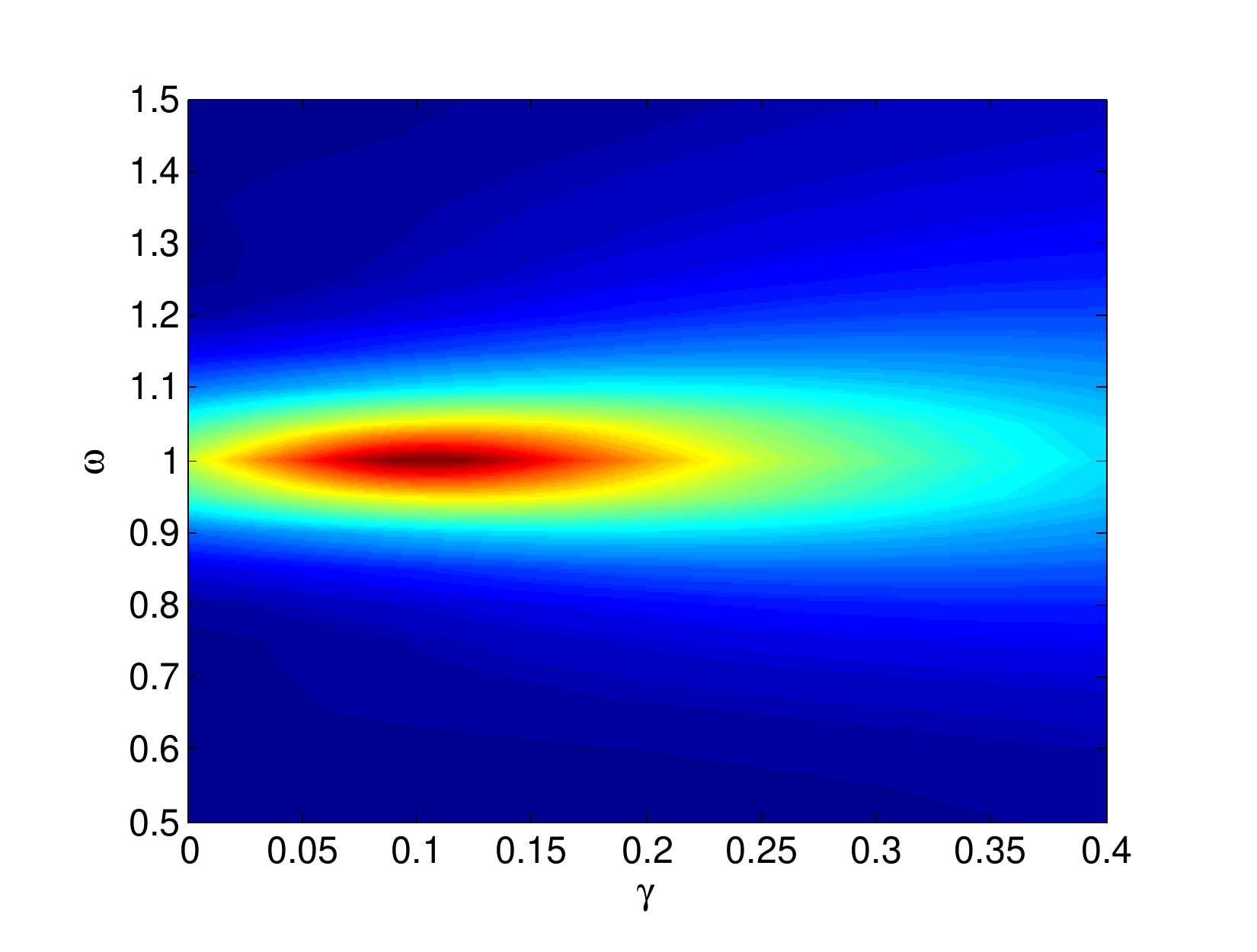}
\caption{Ideal signal and sparse noisy sampling ($N=100$, $T\approx 25$,
$N_e=100$) (left) and log-likelihood function (right) for model system
C with $\omega_y=1$ and $\gamma=0.1$.}  \label{fig:BayesianC}
\end{figure*}

Figs~\ref{fig:BayesianA}--\ref{fig:BayesianC} show that Bayesian
analysis allows us to estimate the model parameters not only for Case A
but for all cases, even if the signal is very noisy, the sampling is
sparse and the measurement and initialization choices are not ideal to
maximize the visibility or signal-to-noise ratio such as
$\theta_M=\tfrac{\pi}{4}$ and $\theta_I=\tfrac{\pi}{3}$.  The squeezed
peaks indicate that the accuracy of the frequency estimation is much
higher than the accuracy of the $\gamma$-estimates.

An additional advantage of the Bayesian estimation does not require
a-priori knowledge of the initialization or measurement angles
$\theta_I$ and $\theta_M$.  Rather, the estimation procedure provides
values for the coefficients of the basis functions, which are related
to the parameters $\theta_I$ and $\theta_M$.

\section{Concluding Discussion}

We have studied the issue of model identifiability and experiment design
for open system dynamics for a dephasing qubit.  From the examples in Sec.
III we can derive some general insights about the limits of identifiability
and the role of experiment design.  Unlike for process tomography, where we
require the ability to prepare the system in many different input states,
and the ability to measure a complete set of observables, we can in general
extract information about both the Hamiltonian and the dephasing parameters
by repeating a single experiment, initializing the system in single fixed
initial state and measuring a fixed basis.  There are certain limitations,
however.  We gain no information about the value of the system parameters
if the initial state is a \emph{stationary state} of the system, or if the
measurement is a \emph{conserved quantity}, although knowledge of the
stationary states or conserved quantities restricts the dynamics and thus
provides indirect information about the system.  Even if the initial state
is not stationary or the measurement is not a conserved quantity, we may
fail to obtain information about the Hamiltonian parameters is if the
operators $H$ and $V$ are orthogonal and $M$ commutes with $H$, for instance.
These limitations also apply to higher dimensional systems, although for
such systems additional restrictions on the identifiability of model
parameters may arise.

If the experiment design is such that the model parameters are identifiable
there are various ways to extract the relevant parameters from a set of
noisy samples (time series), including Fourier analysis, time series analysis
and Bayesian estimation.  Although all of these approaches are in principle
able to provide the required information, Bayesian estimation appears to be
superior to the alternatives, in particular when we are dealing with a limited
number of noisy data points.  One reason the Bayesian estimation is capable
of providing far more accurate estimates for the parameters given the same
input data is that it utilizes information about the structure of the signal
in the form of the choice of the basis functions we are projecting onto.
This allows us to overcome restrictions on the uncertainties of the parameter
estimates imposed by Nyquist's law.

\begin{acknowledgments}
We acknowledge funding from the National Natural Science Foundation
of China (Grant No 60974037).  SGS acknowledges funding from EPSRC
ARF Grant EP/D07192X/1 and Hitachi. We would like to acknowledge
Daniel K. L. Oi. for helpful comments and suggestions.
\end{acknowledgments}

\appendix

\section{Estimating peak frequency and dephasing rate from Fourier spectrum}

Consider a measurement trace of the form
\begin{equation}
   a + b e^{-\gamma t}\sin(\omega_0t),
\end{equation}
which corresponds directly to the expected result for Model A if we set
$a=\cos(\theta_I)\cos(\theta_M)$ and $b=\sin(\theta_I)\sin(\theta_M)$.
If $\theta_I = \theta_M=\pi/2$ then $a=0$ and $b=1$.  A similar analysis
can be done for Models B and C.

The Fourier transform of $u(t)e^{-\gamma t}\sin(\omega_0t)$, where
$u(t)$ is the Heavyside function, is given by $F(\omega) =
\frac{\omega_0}{\omega_0^2+(\gamma+i\omega)^2}$.  We are interested in
its absolute value
\begin{equation}
 |F(\omega)|
 = \frac{\omega_0}{[\gamma^2+(\omega_0-\omega)^2][\gamma^2+(\omega+\omega_0)^2]}
 \end{equation}
Differentiating with respect to $\omega$ and setting the numerator to
$0$ shows that $|F(\omega)|$ has extrema for $\omega
(\omega_0^2-\gamma^2-\omega^2)=0$, and in particular we have a maximum at
$\omega=\sqrt{\omega_0^2-\gamma^2}$ with peak value $(2\gamma)^{-1}$.

Thus, we could in principle estimate both the frequency $\omega_0$ and
dephasing rate $\gamma$ from the peak height $|F|_*$ and position
$\omega_*$, $\gamma=(2 |F|_*)^{-1}$ and
$\omega_0=\sqrt{\omega_*^2+\gamma^2}$, but in practice estimating the
height of the peak is delicate and this approach is usually very
inaccuate.

We can get a better estimate for $\gamma$ using the width of the peak.
Let $\omega_{1,2}$ be the (positive) frequencies for which $|F(\omega)|$
assumes half its maximum or $1/(4\gamma)$.  Then the
full-width-half-maximum $2d$ of $|F(\omega)|$ is $|\omega_2-\omega_1|$
or
\begin{align*}
  d &= \left[\sqrt{\omega_0^2-\gamma^2+2\sqrt{3}\omega_0\gamma}-\sqrt{\omega_0^2-\gamma^2} \right]\\
    &= \left[\sqrt{\omega_*^2+2\sqrt{3} \gamma\sqrt{\omega_*^2+\gamma^2}}-\omega_* \right]
\end{align*}
Given the location $\omega_*$ and half-width $d$ of the peak we can
solve this equation for $\gamma$
\begin{equation}
  \gamma = \frac{1}{6} \sqrt{6 g(\omega_*,d)-18\omega_*^2}
\end{equation}
where $g(\omega_*,d)=\sqrt{9 \omega_*^4 + 12 d^2 \omega_*^2 + 12 d^3
\omega_* + 3 d^4}$.  Thus, in principle we can determine both the
frequency and the dephasing rate by estimating the position and width of
the fourier peak.


\end{document}